\documentclass[letterpaper,11pt]{article}
\pdfoutput=1
\usepackage{jheppub}
\usepackage{amsmath,amssymb,amsfonts,bm,amscd,mathtools}
\usepackage{graphicx}
\usepackage{multirow}
\usepackage{verbatim}
\usepackage{appendix}
\usepackage{fancybox}
\usepackage{array}
\usepackage{url}
\usepackage{float}
\usepackage{subfigure}

\def\CB{{\mathcal B}}

\def\DD{{\mathcal D}}

\def\CF{{\mathcal F}}

\def\CI{{\mathcal I}}

\def\CN{{\mathcal N}}
\def\CO{{\mathcal O}}

\def\CQ{{\mathcal Q}}

\def\be{\begin{equation}}
\def\ee{\end{equation}}
\def\bea{\begin{eqnarray}}
\def\eea{\end{eqnarray}}

\newcommand{\Du}{\sigma}
\newcommand{\Dv}{\tau}
\newcommand{\Dd}{\delta}
\newcommand{\Dl}{\omega}

\newcommand{\za}{{\tt a}}
\newcommand{\zb}{{\tt b}}
\newcommand{\tCC}{{\tilde {\mathcal C}}}
\newcommand{\ct}{{\tt t}}
\newcommand{\st}{{s}}

\title{Constraining Conformal Theories in Large Dimensions}

\author{Abhijit Gadde, Trakshu Sharma}
\affiliation{
Department of Theoretical Physics \\ 
 Tata Institute for Fundamental Research, Mumbai 400005
}

\abstract{
In this paper, we analyze the constraints imposed by unitarity and crossing symmetry on conformal theories in large dimensions. In particular, we show that in a unitary conformal theory in large dimension $D$, the four-point function of identical scalar operators $\phi$ with scaling dimension $\Delta_\phi$ such that $\Delta_\phi/D<3/4$, is necessarily that of the generalized free field theory. This result follows only from crossing symmetry and unitarity. In particular, we do not impose the existence of a conserved spin two operator (stress tensor).
We also present an argument to extend the applicability of this result to a larger range of conformal dimensions, namely to $\Delta_\phi/D<1$. This extension requires some reasonable assumptions about the spectrum of light operators. Together, these results suggest that if there is a non-trivial conformal theory in large dimensions, not necessarily having a stress tensor, then its relevant operators must be exponentially weakly coupled with the rest.
}

\preprint{TIFR/TH/20-6}

\begin{document}
\maketitle
\flushbottom

\section{Motivation}
It is usually believed that there are no non-trivial conformal field theories (CFTs) in greater than six dimensions. This belief stems from thinking of CFTs as infrared fixed points of RG flows that initiate at the free theory. The RG flow is triggered by a relevant or marginally relevant operator about the free theory. The number of relevant and marginal operators decreases with the dimensionality of space-time. In dimensions greater than six free theories do not have any relevant or marginal operators. This leads to the commonly held belief\footnote{
The usual quartic coupling of scalars $(\phi_i\phi_i)^2$ is irrelevant in $D> 4$ dimensions, nevertheless for those dimensions, one can try to find the associated UV fixed point. In \cite{Fei:2014yja}, authors identified this fixed point as a theory of $\phi_i$ and $\sigma$ interacting with $\phi_i\phi_i \sigma+\sigma^3$ potential. Unfortunately, the dimension of operator $\phi_i\phi_i$ goes below the unitary bound for $D>6$.}.

With the discovery of a plethora of the so-called non-Lagrangian superconformal field theories (SCFTs) in the supersymmetric context, one can debate whether a conformal field theory can always be thought of as an endpoint of the RG flow emanating from a free theory. These non-Lagrangian SCFTs do not provide counterexamples in dimension greater than six because SCFTs do not exist in dimensions greater than six. However the reason for this is strictly a kinematical one: non-existence of superconformal algebra (such that supercharges transform in the spinor representation of the Lorentz group) \cite{Kac:1977em, Shnider:1988wh}. Hence the space of SCFTs is a poor diagnostic of the space of CFTs in large dimensions. The question of the existence of CFTs in large dimensions is only meaningful for non-supersymmetric CFTs. Nevertheless, SCFTs yield examples that are perhaps truly non-perturbative in nature. This means we need to analyze the space of CFTs in large dimensions using non-perturbative methods. In this paper, we do so using conformal bootstrap. This approach that started with the work of \cite{Rattazzi:2008pe} has turned out to be very effective in putting numerical constraints on the CFT data, for example, on the critical exponents in $3d$ Ising model \cite{ElShowk:2012ht} as well as obtaining analytical results about the spectrum at large spin \cite{Komargodski:2012ek, Fitzpatrick:2012yx}. See \cite{Poland:2018epd} for a review of the subject and a comprehensive list of references.

If CFTs do exist in large dimensions then using AdS/CFT correspondence they define non-perturbative quantum gravity in large dimensions. If all quantum theories of gravity come from string theory, then this suggests that string theory, albeit strongly coupled,  admits large dimensional AdS vacua. Thus the question of the existence of CFTs in large dimensions is an important one also from the point of view of string theory/quantum gravity. 

Let us now briefly discuss what we mean by a non-trivial CFT. First, note that theories of free massless scalars and free massless fermions are unitary conformal field theories and they exist in arbitrary integer dimension. The theory of massless $(D-2)/2$ forms in $D$ dimension (for even $D$) is also a unitary conformal field theory. There are no other free theories that are conformal. Free scalars, free fermions and free $(D-2)/2$ forms all have stress tensor and hence admit a local coupling to gravity. If we relax the condition of existence of stress tensor then a family of ``generalized free field theories" (GFFTs) can be easily constructed. In GFFT a higher point correlation function is defined as the sum of products of two-point function (which is fixed by conformal symmetry) i.e. as a Wick contraction. Although GFFT can be defined for fields transforming in any representation of the Lorentz group, only the case of scalars is relevant for this paper. For us, GFFTs  (including the genuine free theories that have the stress tensor) are trivial conformal theories.

As we search for non-trivial theories we may either want to impose a) \emph{unitarity} and b) \emph{existence of stress tensor} or not. The physical expectation of having no non-trivial CFTs above a critical dimension is when one imposes both the conditions\footnote{In \cite{Cordova:2019cvf} authors construct $AdS_8$ solution of string theory which would be dual a unitary conformal field theory in $7d$. This would contradict the standard belief that the critical dimension is $6$.}. If we relax either of the two then the existence of a non-trivial theory is plausible. 
Families of non-trivial conformal field theories have been constructed and studied in \cite{Stergiou:2015roa, Gracey:2015xmw, Osborn:2016bev, Guerrieri:2016whh, Brust:2016gjy, Gliozzi:2016ysv, Gromov:2017cja} that exist in high dimensions but are non-unitary.
On the other hand examples of unitary theories but without stress tensor can potentially be constructed by coupling two GFFTs with a relevant operator and flowing down the renormalization group\footnote{Examples of such flow include the double trace deformations studied in \cite{Witten:2001ua, Berkooz:2002ug}. The endpoint of the flow is dual to a change of boundary conditions in the bulk.}\footnote{We thank David Simmons-Duffin for pointing out this possibility to us.}.  
Hence, one would like to conjecture that there are no non-trivial unitary conformal field theories with a stress tensor in sufficiently large dimensions. However, we can constrain the space of conformal theories with certain properties even without requiring a stress tensor.  In particular, we will show that the unitarity and crossing symmetry constrain the four-point function of scalar operators with scaling dimension  $\Delta_\phi/D<1$ to be that of the GFFT in a particular  Lorentzian diamond.

The use of space-time dimension $D$ as an approximation parameter is not new. In the past, it has been used to study quantum gravity in large dimensions \cite{Strominger:1981jg}. In recent years, the large $D$ limit has also been applied to general relativity obtaining a  dramatic simplification in the black-hole dynamics e.g. see \cite{Emparan:2013moa, Bhattacharyya:2015dva}.  

\subsection{Structure of the argument}
In this subsection we give a quick overview of the paper, highlighting the structure of our argument. In large $D$ limit, unitarity forces $\Delta$ also be $\CO(D)$. We define $\Delta=\Dd D$ and also spin as $\ell=\Dl D$. With this scaling, the correlator which is expressed as the sum over conformal blocks can be approximated by an integral over conformal blocks multiplied by OPE coefficient ``density". We argue that this integral is of Laplace type and can be performed by saddle point approximation. The positivity of the OPE coefficient density can be exploited to argue that the real saddle points must lie in the unitary domain. Remarkably it turns out that this condition is incompatible with unitarity for $\Dd_\phi <3/4$ (for $\Dd_\phi<1$ if one assumes sparseness of the low lying spectrum) except for the saddle point that corresponds to the GFFT. 

\subsection*{Outline}
In section \ref{CBlargeD} we motivate the scalings and compute the conformal block in the large $D$ scaling limit. Here the large $D$ solution of \cite{Fitzpatrick:2013sya} plays an important role. We independently check that our solution satisfies the conformal Casimir equation. In section \ref{block-expansion}  we approximate the conformal block expansion by an integral and find the region in the cross-ratio space where the conformal blocks are positive and both s-channel and t-channel OPE are convergent. The consequence of crossing symmetry and unitarity are then analyzed in section \ref{crossing} leading to the main conclusion. In section \ref{discussion}, we summarize our conclusions with outlook. The paper is supplemented with three appendices. Appendix \ref{CB}  gives explicit formulas for conformal blocks in our large $D$ scaling limit. Appendix \ref{saddle-general} illustrates some features of saddle point integrals that are relevant to the discussion in the paper. 
In appendix \ref{solving-crossing} we present a detailed analysis of the constraints of unitarity and crossing.

\section{Conformal blocks at large $D$}\label{CBlargeD}
In this paper we will be concerned with unitary CFTs. The conformal dimension of local operators in bounded from below by unitarity
\bea
\Delta &\geq& \frac{D}{2}-1, \quad\quad\,\,{\rm for}\,\,\ell=0,\nonumber\\
\Delta &\geq& \ell +D-2, \quad{\rm for}\,\,\ell\neq0.
\eea
In any large $D$ limit, a unitary conformal field theory has local operators with dimensions that scale linearly with $D$. We take $\Delta=\Dd D$ with $\Dd$ fixed. The eigenvalue of the conformal Casimir for the conformal multiplet with primary of dimension $\Delta$ and spin $\ell$ is $\Delta(\Delta-D)+l(l+D-2))$. In order for the spin to contribute to conformal block we also take spin to scale linearly with $D$, $\ell=\Dl D$ with $\Dl$ fixed. Conformal blocks with finite spin can be obtained by setting $\Dl$ to be $\CO(D^{-1})$.

The conformal blocks satisfy the conformal Casimir equation. In the large $D$ limit, this equation can be separated and solved. This was done in \cite{Fitzpatrick:2013sya}. We have reproduced their result below
\bea
\CF_{\Delta,\ell}(u,v)&=&\frac{2^{\Delta+\ell}}{\sqrt{y_+-y_-}} A_{\Delta}(y_+)A_{1-\ell}(y_-)\qquad {\rm where}\quad y_{\pm}=\frac{u}{(1\pm \sqrt v)^2},\nonumber\\
A_{x}(y)&=&y^{x/2}\,\,_2F_1\Big(\frac{x-1}{2},\frac{x}{2},x-\frac{D}{2}+1;y\Big).
\eea
Here $(u,v)$ are the standard conformal cross-ratios defined as $u=x_{12}^2x_{34}^2/x_{13}^2x_{24}^2$ and $v=x_{14}^2x_{23}^2/x_{13}^2x_{24}^2$.
Above approximation is valid for $y_+-y_- \gg 1/D$ or equivalently $v\gg 1/D^2$. To compute the blocks in the scaling limit $\Delta=\Dd D$ and $\ell = \Dl D$ we need to approximate the hypergeometric function at large values of parameters. This can be done by expressing the hypergeometric function in the Euler integral form and performing the integral using the saddle point. The result is of the form
\bea\label{laplace}
&&\CF_{D\Dd,D\Dl}(u,v) \xrightarrow{D\to \infty} \CN_{\Dd,\Dl}(u,v)(1+\CO(\frac1D)) \nonumber\\
&&\CN_{\Dd,\Dl}(u,v) =  \Big(f_{\Dd}(u,v)e^{D \, g_{\Dd}(u,v)}\Big) \Big(f_{\Dl}(u,v)e^{D \, g_{\Dl}(u,v)}\Big)
\eea
where $f_{\Dd},f_{\Dl}$ and $g_{\Dd},g_{\Dl}$ are complicated functions of the arguments and labels. 
We do not give their explicit form here as it does not offer much insight. It is given in appendix \ref{CB}. 

Although we have not checked explicitly, we believe the perturbative corrections in $1/D$  to $\CN_{\Dd,\Dl}$ form a convergent series and the non-perturbative corrections i.e. $e^{-D}$ corrections are absent. This means a given block contributes a specific cross-ratio dependent exponential piece 
 $e^{D  \, (g_{\Dd}(u,v)+g_{\Dl}(u,v))}$ to the correlator. If the two blocks have either $\Dd$ or $\Dl$ that is $\CO(1)$ different then their exponential contributions are distinct. However, if one considers blocks for $(\Dd',\Dl')$ and $(\Dd,\Dl)$ such that $\Dd'=\Dd+\CO(\frac1D)$ and $\Dl'=\Dl+\CO(\frac1D)$ then their exponential contribution can not be distinguished. This is because, 
 \be
 e^{D g_{\Dd',\Dl'}(u,v)} = c_{\Dd,\Dl}(u,v) e^{D g_{\Dd,\Dl}(u,v)}, \qquad \text{ for some} \quad c_{\Dd,\Dl}(u,v)\sim \CO(1).
 \ee
 These considerations are important as our arguments will essentially involve matching distinct exponential contributions of  blocks.

 The expression for the block simplifies if we look at the dependence in a small neighborhood of size $1/D$ around a certain point. We do this by substituting $(u,v)\to (\za^2 \, e^{\Du/D}, \zb^2 \,e^{\Dv/D})$ and focusing on the dependence on $(\Du,\Dv)$. 
\bea\label{largeDblock}
\CF_{D\Dd,D\Dl}(u,v) \xrightarrow{D\to \infty} &&\CN_{\Dd,\Dl}(\za^2,\zb^2)\,\,\CB_{\Dd,\Dl;\za,\zb}(\Du,\Dv) \\
\CB_{\Dd,\Dl;\za,\zb}(\Du,\Dv) =&& e^{k_{+}(\za,\zb,\Dd)\big((1+\frac1\zb) \Du-\Dv\big)} e^{k_{-}(\za,-\zb,1+\Dl)\big((1-\frac1\zb) \Du-\Dv\big)}\nonumber\\
{\rm where}\quad k_{\pm }(\za,\zb,x)=&& \frac{\zb(1+\zb)}{4((1+\zb)^2-\za^2)}\Big(1\pm\sqrt{1+4x(x-1)\big(1-\frac{\za^2}{(1+\zb)^2}\big)}\Big).\nonumber
\eea
This is the large $D$ approximation to the conformal blocks that we will work with in the rest of the paper. When the block is thought of as a function of $(\Du,\Dv)$, $\CN_{\Dd,\Dl}$ is interpreted as the ``normalization". The $(\Du,\Dv)$ dependence is the imprint of the function $g_{\Dd,\Dl}(u,v)$ in an $\CO(1/D)$ neighborhood. Hence matching of exponential contributions is tantamount to matching the $(\Du,\Dv)$ dependence in the exponent. This is what we will do in imposing crossing symmetry.

The $(\Du,\Dv)$ dependence of the conformal block in the large $D$ limit can be verified by checking that it satisfies the conformal Casimir equation in the scaling limit described above.  Recall that the conformal Casimir equation is
\be
(L_{(1)}+L_{(2)})^2 {\CF}_{\Delta,\ell}=C_{\Delta,\ell} \CF_{\Delta,\ell}.
\ee
The eigenvalue $C_{\Delta,\ell}=\Delta(\Delta-D)+\ell(\ell+D-2)$. In terms of the conformal cross-ratios $(u,v)$ this reduces to the following coupled second-order differential equation. 
\bea
&&\Big(-((1-v)^2-u(1+v))\partial_v v\partial_v-(1-u+v)u\partial_u u\partial_u \nonumber\\
&&+2(1+u-v)uv\partial_u\partial_v+Du\partial_u +\frac{C_{\Delta,\ell}}{2}\Big) \CF_{\Delta,\ell}(u,v)=0.
\eea 
After substituting $\Delta=D\Dd, \ell= D\Dl$ and $(u,v)= (\za^2 \, e^{\Du/D}, \zb^2 \,e^{\Dv/D})$ and taking the large $D$ limit we get,
\be
\Big(-\frac{((1-\zb^2)^2-\za^2(1+\zb^2))}{\zb^2} \partial_\Dv^2-(1-\za^2+\zb^2)\partial_\Du^2+2(1+\za^2-\zb^2)\partial_\Du\partial_\Dv+\partial_\Du+\frac12c_{\Dd,\Dl}\Big) \CF=0,
\ee
where  $c_{\Dd,\Dl}=(\Dd(\Dd-1)+\Dl(\Dl+1))$. The variables $(\za,\zb)$ are simply constants.  It is straightforward to check that the large $D$ block \eqref{largeDblock} satisfies this equation.

To see why the conformal block is dominated by a monomial of $( e^{\Du/D}, e^{\Dv/D})$, it is instructive to consider the series expansion of a  scalar conformal block that is known in arbitrary dimension. 
\be\label{scalar-series}
\CF_{\Delta,0}=\sum_{m,n=0}^{\infty}\frac{(\Delta/2)_n^2(\Delta/2)_{n+m}^2}{(\Delta+1-D/2)_n(\Delta)_{2n+m}}\frac{u^{\frac{\Delta}{2}+n}}{n!}\frac{(1-v)^{m}}{m!}, \qquad (x)_n\equiv \Gamma(x+n)/\Gamma(x).
\ee
In our large $D$ scaling limit the sum over descendants becomes a saddle point integral. This is performed in appendix \ref{CB}. At a given value of $(\za,\zb)$, a single descendent with dimension $k_{+}(\za,\zb,\Dd)(1+\frac1\zb)D$ dominates sum and hence the $(\Du,\Dv)$ dependence of the conformal block takes the simple form \eqref{largeDblock}.

\section{Conformal block expansion at large $D$}\label{block-expansion}
In this paper we will be concerned with four point function of identical scalar operators of dimension $\Delta_\phi=D\Dd_\phi$. The stripped correlation function \emph{i.e.} $\langle\phi(x_1)\ldots\phi(x_4)\rangle (x_{12}^{2} x_{34}^2)^{\Delta_\phi}$ is only a function of cross-ratios, $G(u,v)$.  It is expanded in terms of s-channel conformal blocks as follows
\be
G(u,v)=1+\sum_{\frac{D-2}{2}\leq \Delta < D-2}  \tCC_{\Delta,0} \CF_{\Delta,0}(u,v)+\sum_{\ell \geq 0}\sum_{\Delta \geq D-2+\ell}  \Big(1+(-1)^\ell\Big)\tCC_{\Delta,\ell} \CF_{\Delta,\ell}(u,v).
\ee
The range of the sum is controlled by unitarity. We have divided the sum into two parts, the first part is supported on the range of $\Delta$ for scalar operators   $(D-2)/2 \leq \Delta< D-2$ and the second part is the range  $(D-2) \leq \Delta-\ell$ for all $\ell$. This is the unitary domain. The first part of the sum is supported over what we call the $\DD_1$ domain and the second part, over the $\DD_2$ domain. The factor of $1+(-1)^\ell$ picks only the contribution of the even spins as desired.

\begin{figure}[h]
    \centering
    {\includegraphics[width=0.5\textwidth]{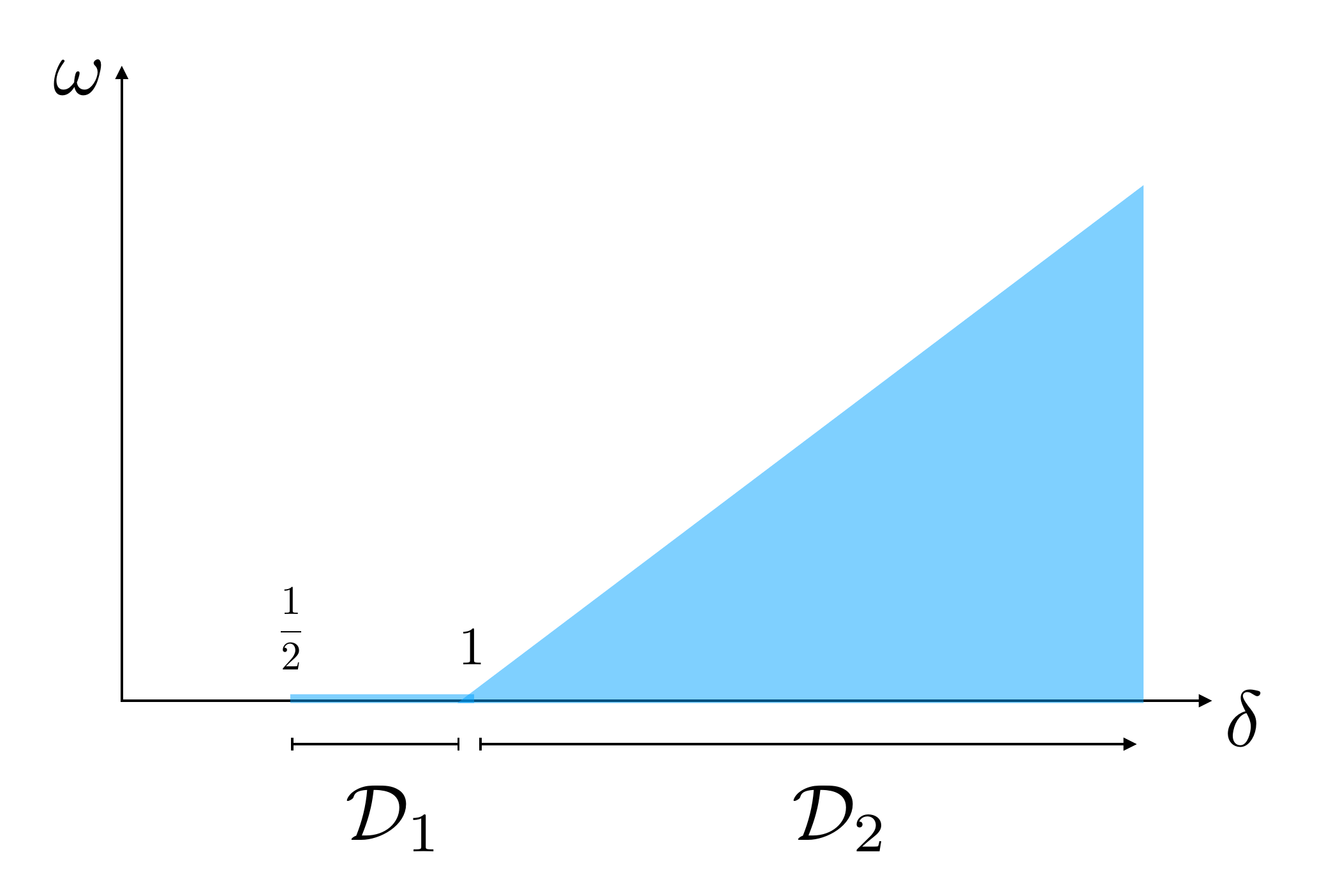} } \hspace{.1cm}
    \caption{Unitary domain in $(\Dd,\Dl)$ plane. We have shown the two pieces, $\DD_1$ and $\DD_2$. }\label{unitary-domain}
\end{figure}

For now, let's replace the factor $1+(-1)^\ell$ by $1$ in taking the large $D$ limit. We will account for this error in section \ref{u-symmetry}.
 In the large $D$ limit, $\Delta=D\Dd$ and $\ell=D\Dl$, it is convenient to replace the sums over $\Delta$ and $\ell$ to integrals over $\Dd$ and $\Dl$ and replace the OPE coefficients by OPE coefficient density  $C_{\Dd,\Dl}=\tCC_{D\Dd,D\Dl}$. 
The OPE density consists of a collection of Dirac delta functions at $(\Dd,\Dl)$ of all operators appearing in the OPE with then strength given by the OPE coefficient.
 At leading order, 
\bea\label{largeDcorrelator}
G(\za^2 e^{\Du/D},\zb^2 e^{\Dv/D})&=&1+D \int_{\DD_1} d\Dd \,  \Big(C_{\Dd,0} \CN_{\Dd,0}(\za^2,\zb^2)\Big)\,\, \CB_{\Dd,0;\za,\zb}(\Du,\Dv)\nonumber\\
&+& D^2 \int_{\DD_2} d\Dd d\Dl \,  \Big(C_{\Dd,\Dl} \CN_{\Dd,\Dl}(\za^2,\zb^2)\Big)\,\, \CB_{\Dd,\Dl;\za,\zb}(\Du,\Dv).
\eea 
The explicit expression for $\CB$ is given in equation \eqref{largeDblock}. We have grouped the terms independent of $(\Du,\Dv)$ in the brackets.
In the large $D$ limit, the domains of integration are $\DD_1=\{(\Dd,\Dl):\Dl=0,1/2\leq \Dd<1\}$ and $\DD_2=\{(\Dd,\Dl):\Dl>0,\Dd\geq \Dl+1\}$.  In terms of $(\Dd,\Dl)$ coordinates we have graphically presented the unitary domain in figure \ref{unitary-domain}.

Consider the unlikely situation where the Dirac delta functions in the OPE density are $\CO(1)$ spaced either in $\Dd$ or in $\Dl$. Then as argued below equation \eqref{laplace}, each operator in the block expansion yields a distinct exponential cross-ratio dependence to the correlator. Of course, more likely the OPE density consists of closely spaced i.e. $\CO(1/D)$ spaced Dirac delta functions. This motivates the definition of  a ``smeared OPE density" $C_{\Dd,\Dl}^{\epsilon}$. We define it to be the OPE density $C_{\Dd,\Dl}$ averaged over squares of size $\epsilon\sim \CO(1/D)$ in $(\Dd,\Dl)$ space. We will pick $\epsilon$ to be the smallest such that $\log(C_{\Dd,\Dl}^{\epsilon})$ is smooth to leading order in $D$ at the scale of $1/D$ in $(\Dd,\Dl)$ space. 
This means $\log(C_{\Dd,\Dl}^{\epsilon})$ is piecewise smooth at scale of $\CO(1)$ in $(\Dd,\Dl)$ to leading order in $D$. Note that the unitarity condition that the OPE coefficient $\tCC_{\Delta,\ell}$ is positive implies that $\log(C_{\Dd,\Dl}^\epsilon)$ to leading order in $D$ is real. 

The reason we demand the smoothness for logarithm and not the OPE density itself will become clear soon. Note that our condition is weaker than the smoothness for the function itself.
Let us replace the very jagged function $C_{\Dd,\Dl}$ with the function with piecewise smooth logarithm $C_{\Dd,\Dl}^\epsilon$. We will quantify the error associated with this replacement shortly.

 Now we are in the position to see why the analysis of \eqref{largeDcorrelator} is viable. The key property is the exponential dependence in $D$ in \eqref{laplace}. This makes the integral in \eqref{largeDcorrelator} in each piecewise smooth region of Laplace type. The feature of such integrals is that they are dominated by discrete points. These points could either be the saddle points of the integrand or be points on the boundary of the region. 
We call all such points ``locally dominant points". We direct the interested reader to appendix \ref{saddle-general} for more discussion of the generalities of such integrals.

In addition to the conformal block $\CN_{\Dd,\Dl}$ having $e^{D\ldots}$ dependence, we will now argue that generically the smeared OPE coefficient density $C^\epsilon_{\Dd,\Dl}$ also has $e^{D\ldots}$ dependence.  
 If the OPE coefficients go as $e^{D^{\alpha}\ldots}$ where $\alpha>1$ then the saddle point will be determined only by the OPE coefficient and the large $D$ correlator will simply be a single conformal block evaluated at that saddle point $(\Dd_*,\Dl_*)$ of $C^\epsilon_{\Dd,\Dl}$. On the other hand, if $\alpha <1$ then the saddle point will be determined completely kinematically by the $e^{D\ldots}$ factor in the conformal block $\CN_{\Dd,\Dl}$.  Most general saddle points are obtained if OPE coefficient density also goes as $e^{D\ldots}$. In the rest of the paper, we assume that to be the case\footnote{This is indeed the case for GFFT. See section \ref{mft}.}. All in all, the position of the saddle point depends on the factor inside the bracket. This is precisely the part that is independent of $(\Du,\Dv)$. That is what makes the correlator simpler to compute in a small neighborhood of $(\za^2,\zb^2)$ of size $1/D$.

The position of the saddle point, more generally of the point that dominates the integral, is inside the integration range, in particular in the unitary domain $\DD_1 \cup \DD_2$, if the integrand does not have a rapidly oscillating phase \emph{i.e.} of the type $e^{iD\ldots}$ where $\ldots$ stand for a function of $(\Dd,\Dl)$. The rapid oscillations are precisely what we dropped when we replaced $C_{\Dd,\Dl}\to C_{\Dd,\Dl}^\epsilon$. The contribution from such rapid oscillations is a subleading saddle point that will lie in the complex domain of $(\Dd,\Dl)$. Hence, the error associated with this replacement is an exponentially subdominant one. We ignore this error as we are interested in constraining only the exponentially dominant part of the correlator.

The conformal block $\CN_{\Dd,\Dl}(\za^2,\zb^2)$ could also be rapidly oscillating.  This would give rise to locally dominant points that are outside the integration range. 
To avoid such possibility then we need to consider only that region of cross-ratio space where all conformal blocks are real and non-negative\footnote{Strictly speaking, a region where conformal blocks have a constant phase will also do but it turns out to be simpler to find the region where they are non-negative.} with no rapid oscillations\footnote{Otherwise these rapid oscillations could interfere constructively with the rapid oscillations of the OPE density that we averaged over to give exponentially dominant contributions.}. In such a region, the integral \eqref{largeDcorrelator} is dominated by discrete points which lie in the integration range i.e. the unitary domain $\DD_1 \cup \DD_2$. Moreover, each locally dominant point yields a distinct exponential cross-ratio dependence. As we move in the cross-ratio space, these points move in a continuous way. If a saddle point moves to the boundary of the smooth region, it dominates the integral as a boundary point.
This seemingly unremarkable constraint on the location and movement of the dominant points will turn out to be extraordinarily powerful when paired with crossing symmetry. In the next two subsections,  we will look for such region in the cross-ratio space where all s-channel, as well as all t-channel conformal blocks, are positive. Needless to say that we will also demand that the s-channel and t-channel expansions be convergent in this region.

\subsection{Positivity of conformal blocks}
It is known that all the unitary conformal blocks have positive coefficients when expanded in terms of variables $(z,\bar z)$, where $u=z\bar z$ and $v=(1-z)(1-\bar z)$. This was first shown in \cite{Fitzpatrick:2012yx} and then in \cite{Hartman:2015lfa}, in both by essentially using positivity of the norm of certain descendent states. Because, $a_{m,n}\geq 0$ where
\be\label{positive}
\CF_{\Delta,\ell}(u,v)=z^{\frac{\Delta+\ell}{2}}{\bar z}^{\frac{\Delta-\ell}{2}}\sum_{m,n\geq 0}^{\infty} a_{m,n}\,\, z^m {\bar z}^n,
\ee
the conformal block is positive for real $z,{\bar z}\geq 0$. In our analysis of the crossing equation, we would also like to demand the t-channel conformal block to be positive. This forces $1-z,1-\bar z \geq 0 $. To summarize,  both the s-channel and the t-channel conformal blocks are positive in the diamond  $0\leq z, \bar z \leq 1$ with real $(z,\bar z)$. The conformal blocks are explicitly computed in appendix \ref{CB} and indeed they are positive and with no rapid oscillations.

\subsection{Convergence of the OPE}
In addition to having positive s-channel and t-channel conformal blocks, we are also interested in having a convergent OPE expansion both in s-channel and in t-channel. The positivity of the blocks discussed in the above subsection also helps determine the regions of convergence of the OPE. In fact, both in \cite{Fitzpatrick:2012yx} and in \cite{Hartman:2015lfa}, the positivity of the blocks was used to do precisely that. From state operator correspondence it follows that the s-channel OPE is convergent over the entire range of cross-ratios in the Euclidean regime except for the half-line $z=\bar z \geq 1$. Similarly, the t-channel OPE is convergent in the  Euclidean regime except for the half-line $z=\bar z \leq 0$. In particular, the s-channel and the t-channel OPE are convergent for $0< z=\bar z <1$. From here, the positivity of the conformal block expansion coefficients in $z,\bar z$ helps us deduce that both the s-channel and t-channel OPEs are convergent in the diamond of interest  $0\leq z, \bar z \leq 1$. This is as follows, let $z < \bar z$ in the diamond,
\be
F_{\Delta, \ell}(z,\bar z) < F_{\Delta, \ell}(\bar z, \bar z)
\ee
This is because of the coefficients $a_{m,n}$ appearing in the expansion \eqref{positive} are positive. This means,
\be
\sum_{\Delta,\ell} \tCC_{\Delta,\ell} F_{\Delta, \ell}(z,\bar z) < \sum_{\Delta,\ell} \tCC_{\Delta,\ell} F_{\Delta, \ell}(\bar z,\bar z).
\ee
Again, this is because the OPE coefficients $\tCC_{\Delta,\ell}$ are positive. The right-hand side is simply the s-channel expansion in the range $0< z=\bar z <1$ which makes the sum convergent, implying that the left-hand side is also convergent. The same argument can be repeated if $z > \bar z$. Even the convergence of the OPE in the t-channel in the region $0< z,\bar z <1$  follows from this argument in the same way. Hence, from now on we will totally confine ourselves in the region $0< z,\bar z <1$, or \emph{the diamond} for short. In terms of $(\za,\zb)$, the diamond is $\za,\zb>0, \za+\zb<1$.

\subsection{u-symmetry}\label{u-symmetry}
In this section we account for the error that we made near equation  \eqref{largeDcorrelator} of summing over all spins rather than summing over only even spins.
\be
\sum_{\ell\geq 0, {\rm even}} \sum_{\Delta} \tCC_{\Delta,\ell} \CF_{\Delta,\ell}(u,v)=\sum_{\ell\geq 0, {\rm all}} \sum_{\Delta} \tCC_{\Delta,\ell} \CF_{\Delta,\ell}(u,v)+ \sum_{\ell\geq 0, {\rm all}} \sum_{\Delta} (-1)^\ell \tCC_{\Delta,\ell} \CF_{\Delta,\ell}(u,v)
\ee 
Let us call the first and second term on the right hand side $\ct_1$ and $\ct_2$ respectively. As the OPE coefficients are positive and the conformal blocks are positive in the diamond , it is clear that $\ct_1 >\ct_2$ in the diamond.

Also note that the conformal blocks are u-symmetric in the following way,
\be
(-1)^\ell \CF_{\Delta,\ell}(u,v) = \CF_{\Delta,\ell}(\frac{u}{v},\frac{1}{v}).
\ee
This makes $\ct_2$ the u-symmetric image of $\ct_1$. Written as a sum of $\ct_1$ and $\ct_2$, the correlator is manifestly u-invariant. When we take the large $D$ limit, both  $\ct_1$ and $\ct_2$ are saddle point integrals. The $\ct_1$ integral is the one in equation \eqref{largeDcorrelator}. The $\ct_2$ is the same integral except for an additional insertion of $e^{i\pi \Dl D}$. In the diamond, the saddle point of the $\ct_1$ integral lies in the unitarity domain $\DD_1\cup \DD_2$ but the saddle point for $\ct_2$ will generically be elsewhere in the complexified $(\Dd,\Dl)$ plane. Due to rapid phase oscillations in $\Dl$, $\ct_2$ is exponentially smaller than $\ct_1$. This is the same reason why rapid oscillations in $\Dd$  give rise to exponentially smaller contributions.
In the absence of u-symmetry, we will not be able to say anything about the location of this saddle point. However, we know that the $\ct_2$ saddle point is the u-symmetric image of the $\ct_1$ saddle point.

\section{Solving Crossing symmetry}\label{crossing}
The crossing  equation is,
\be
1+\sum_{\DD_1\cup \DD_2} \Big(1+(-1)^\ell\Big) \tCC_{\Delta,\ell} \CF_{\Delta,\ell}(u,v)=\Big(\frac{u}{v}\Big)^{\Delta_\phi} \Big(1+\sum_{\DD_1\cup \DD_2} \Big(1+(-1)^\ell\Big) \tCC_{\Delta,\ell} \CF_{\Delta,\ell}(v,u)\Big).
\ee
In the large $D$ limit, the sum over conformal blocks is approximated as integrals\footnote{We have suppressed the factors of $D$ outside the integrals for compactness of the expression.}, 
\bea\label{block-int}
1&+&\int_{\DD_1 \cup \DD_2} d\Dd d\Dl \Big(1+e^{i\pi \Dl D}\Big)  \Big(C^\epsilon_{\Dd,\Dl} \CN_{\Dd,\Dl}(\za^2,\zb^2)\Big)\,\, \CB_{\Dd,\Dl;\za,\zb}(\Du,\Dv)  \\
&=&\Big(\frac{\za^2}{\zb^2}\Big)^{\Dd_\phi D} e^{(\Du-\Dv)\Dd_\phi}\Big(1+\int_{\DD_1 \cup \DD_2} d\Dd d\Dl \Big(1+e^{i\pi \Dl D}\Big)  \Big(C^\epsilon_{\Dd,\Dl} \CN_{\Dd,\Dl}(\zb^2,\za^2)\Big)\,\, \CB_{\Dd,\Dl;\zb,\za}(\Dv,\Du)\Big).\nonumber
\eea
We have naturally defined $\Delta_\phi \equiv \Dd_\phi D$. The explicit expression for $\CB$ is given in equation \eqref{largeDblock}. As discussed in section \ref{u-symmetry}, the integrals on both sides are split into two terms $\ct_1$ and $\ct_2$. The  term $\ct_1$ comes from the $1$ and the term $\ct_2$ comes from $e^{i\pi \Dl D}$ in $(1+e^{i\pi \Dl D})$ respectively.

Every correlator of identical operators has a universal contribution, namely the contribution from the identity operator. 
As the contribution of the universal saddle must be crossing dual to identity operator,  it is perhaps not surprising that the universal saddle fixes the OPE density and its contribution to the correlator to be that of the GFFT.
We show this explicitly in section \ref{mft}. For now let us assume that there exists a pair of points $(\Dd_s^*,\Dl_s^*)$ and $(\Dd_t^*,\Dl_t^*)$ in the s-channel and t-channel expansion respectively that dominate over this universal part for some $(\za,\zb)$. If we find that the existence of such a saddle is inconsistent with unitarity and crossing symmetry then we must conclude that the correlator is given by $1$ and its universal dual saddle i.e. by that of the GFFT.  For $\Dd_\phi<1$, we will show that it is indeed the case.

\begin{figure}[h]
    \centering
    \subfigure[No overlap]{\includegraphics[width=0.31\textwidth]{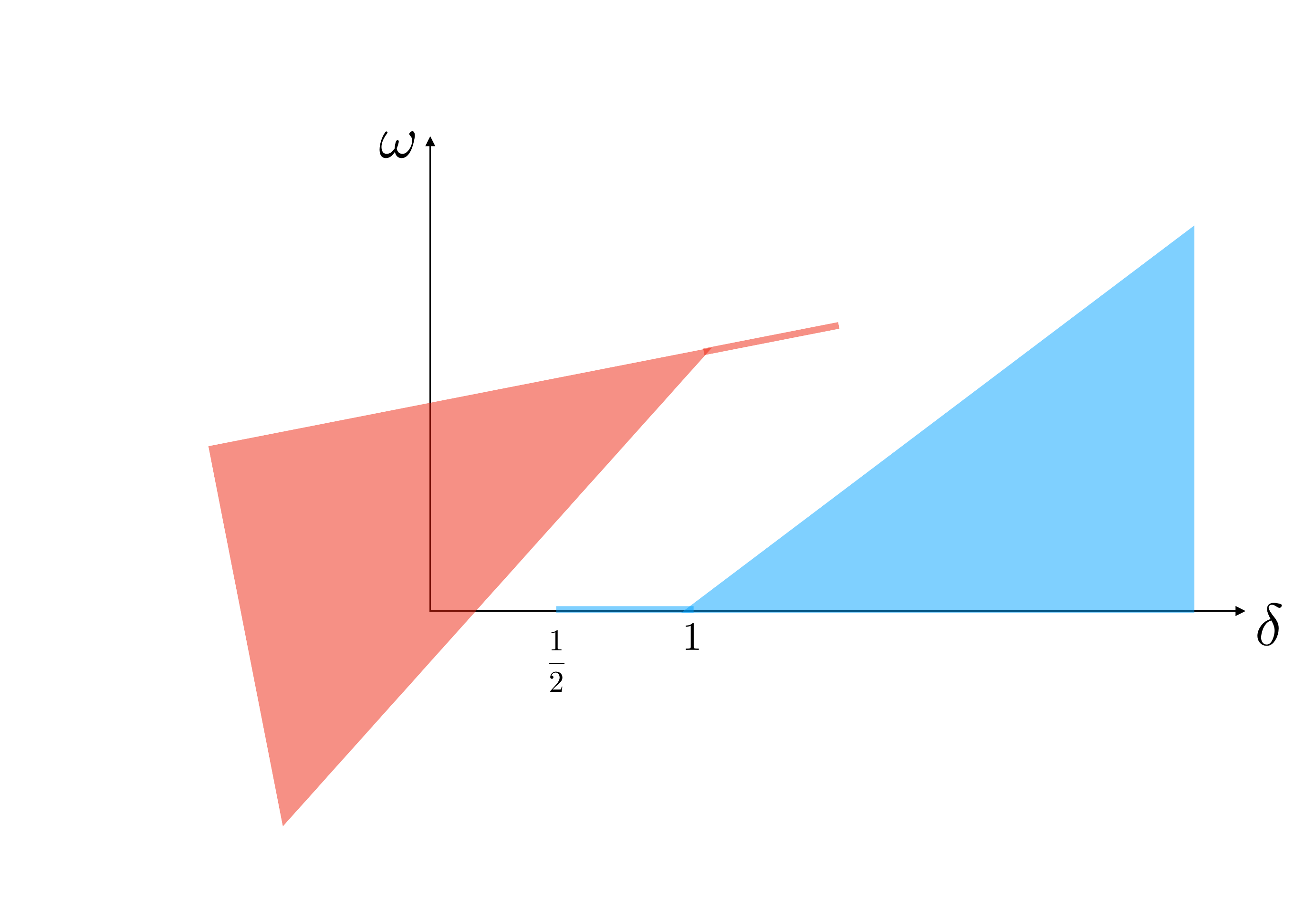} } \hspace{.1cm}
    \subfigure[Type I: $\DD_1^s \cap \DD_1^t$]{\includegraphics[width=0.31\textwidth]{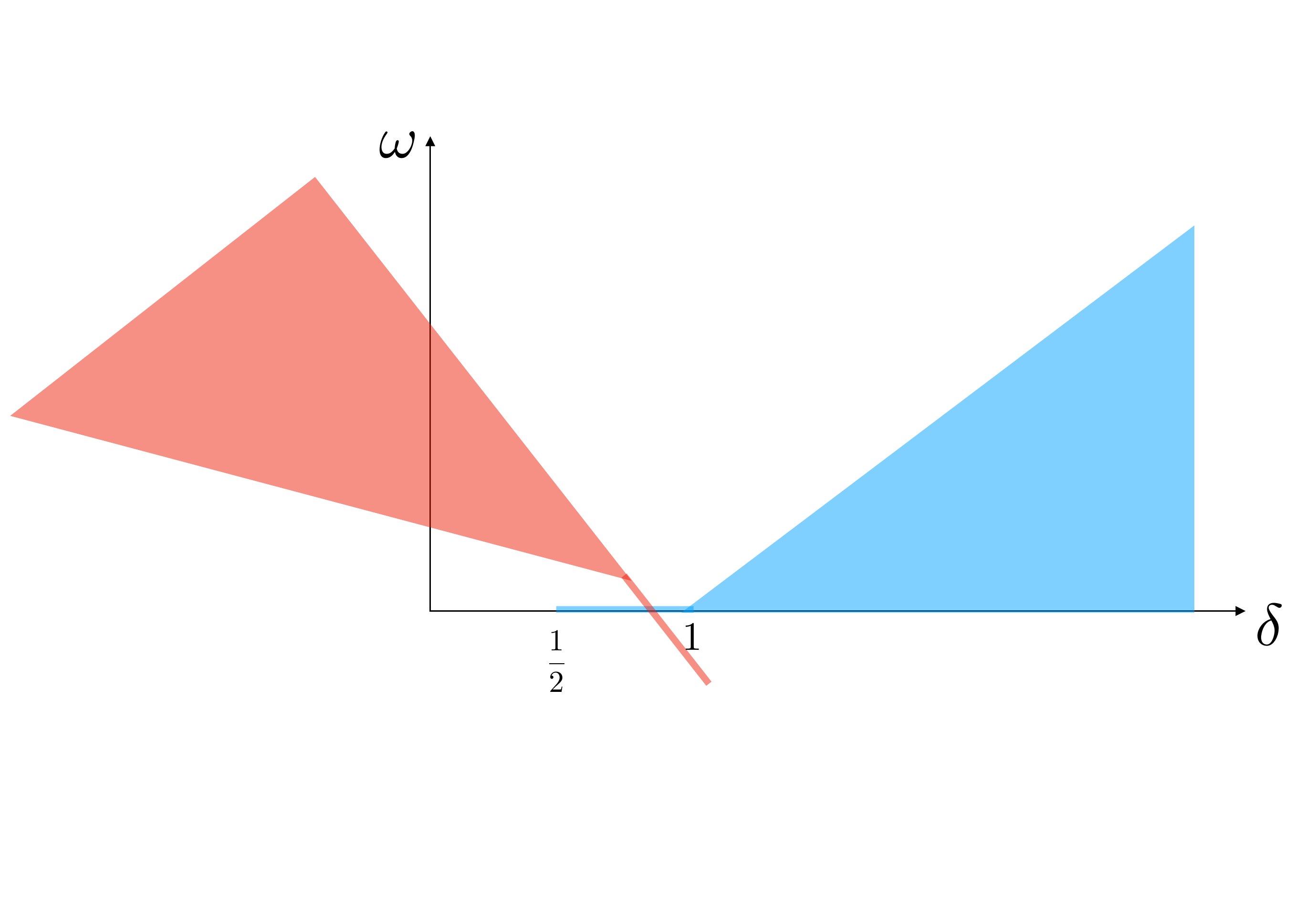} } \hspace{.1cm}
    \subfigure[Type II: $\DD_2^s \cap \DD_1^t$]{\includegraphics[width=0.31\textwidth]{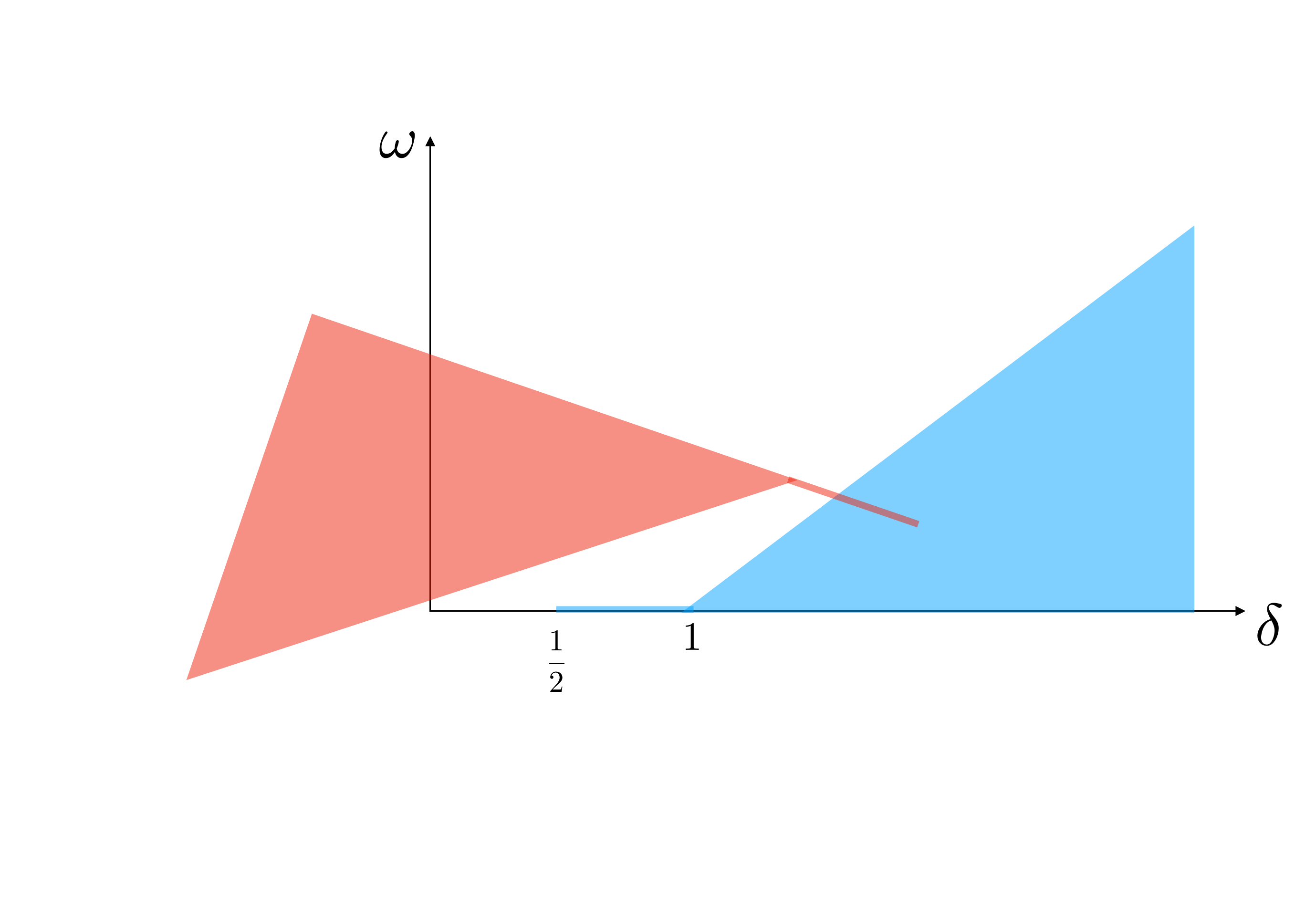}  }\hspace{.1cm}
    \subfigure[Type III: $\DD_1^s \cap \DD_2^t$]{\includegraphics[width=0.31\textwidth]{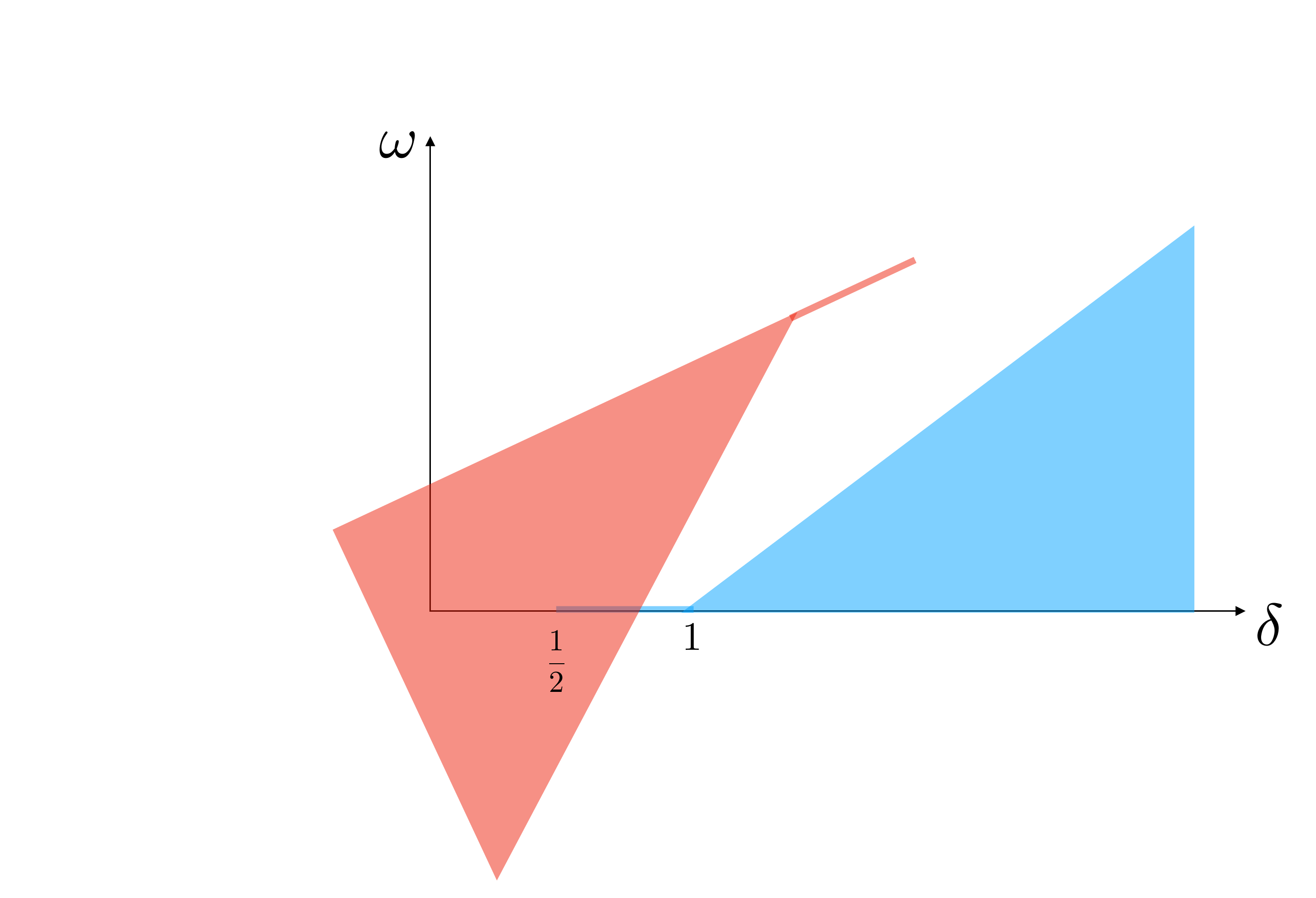}  }\hspace{.1cm}
    \subfigure[Type IV: $\DD_2^s \cap \DD_2^t$]{\includegraphics[width=0.31\textwidth]{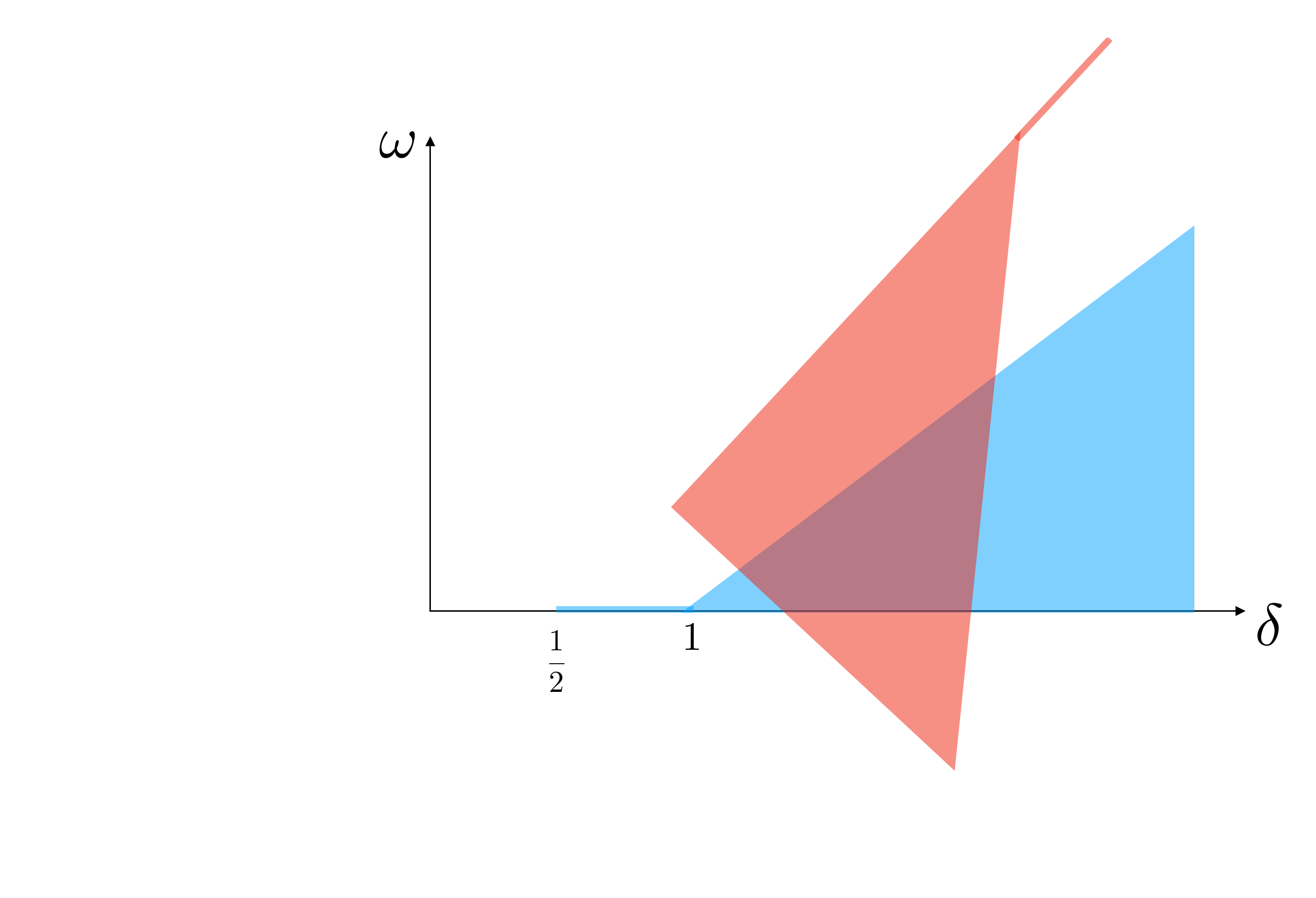}  }\hspace{.1cm}
    \subfigure[Types II + III + IV]{\includegraphics[width=0.31    \textwidth]{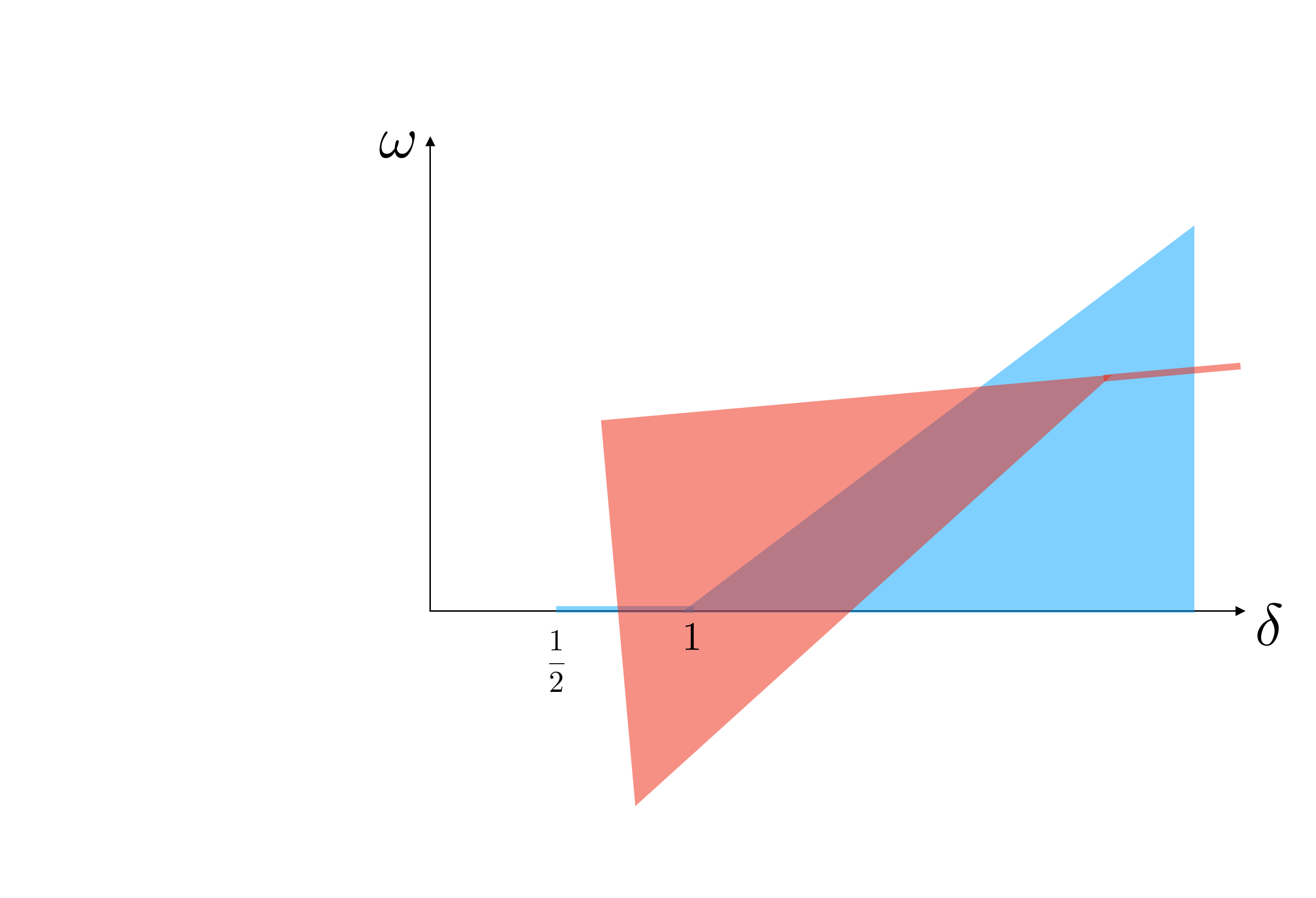}  }
    \caption{In these figures we have denoted the unitary domain in the s-channel in the $(\Dd_1^*,\Dl_1^*)$ plane by blue color and have superimposed the image of the unitary domain in the t-channel under the map \eqref{crossing-saddle} in red. In (a) we do not have any overlap. In such a scenario there are no solution to the crossing equation compatible with unitarity. In subfigures (b), (c), (d), (e) we have shown possible elementary overlaps of $\DD_1$ and $\DD_2$ parts of the unitary domain in the s and t channel. Of course, combinations of such overlaps is also a possibility. For example in (f), overlaps  types II, III and IV occur.}\label{overlaps}
\end{figure}

Let's proceed to a proof by contradiction.   
Let the points $(\Dd_s^*,\Dl_s^*)$ and $(\Dd_t^*,\Dl_t^*)$ in the s-channel and t-channel expansion respectively be globally dominant points at $(\za,\zb)$. 
Both these contributions must necessarily come from $\ct_1$ part of the integral (because $\ct_1$ is exponentially dominant over $\ct_2$) and hence lie in the unitary domain $\DD_1\cup \DD_2$. Focusing only on the $(\Du,\Dv)$ dependent part in exponent on both sides,
\be
\CB_{\Dd_s^*,\Dl_s^*;\za,\zb}(\Du,\Dv) = \#\, e^{(\Du-\Dv)\Dd_\phi}  \CB_{\Dd_t^*,\Dl_t^*;\zb,\za}(\Dv,\Du),
\ee
where $\#$ is some constant that is independent of $(\Du,\Dv)$ (but does depend on $(\za,\zb)$).  Matching the coefficients of $\Du$ and $\Dv$ in the exponent,  we get
\bea\label{crossing-saddle}
&&k_{t+}=\frac{\za-1}{2\zb}(k_{s+}-k_{s-})-\frac12(k_{s+}+k_{s-}-\Dd_\phi),\\
&&k_{t-}=\frac{-\za-1}{2\zb}(k_{s+}-k_{s-})-\frac12(k_{s+}+k_{s-}-\Dd_\phi),\qquad {\rm where}\nonumber\\
\nonumber \\
&&k_{s+}\equiv k_{+}(\za,\zb,\Dd_s^*),\, k_{s-}\equiv k_{-}(\za,-\zb,1+\Dl_s^*),\, k_{t+}\equiv k_{+}(\zb,\za,\Dd_t^*),\, k_{t-}\equiv k_{-}(\zb,-\za,1+\Dl_t^*).\nonumber
\eea
This equation expresses the dominant point in the t-channel $(\Dd_t^*,\Dl_t^*)$ in terms of the dominant point in the s-channel $(\Dd_s^*,\Dl_s^*)$. As we move around the cross-ratio space, then these points may cease to be globally dominant but continue to be locally dominant and will continue to exist in their respective unitary domain. We ask if this is possible i.e.  we ask if a pair of the s-channel and t-channel  points that satisfies equation \eqref{crossing-saddle} exists such that both of them lie in the unitary domain $\DD_1\cup \DD_2$. In other words, we ask if the image of the unitary domain $\DD_1^t \cup \DD_2 ^t$ for the t-channel locally dominant point overlaps with the unitary domain $\DD_1^s\cup \DD_2^s$ for the s-channel locally dominant point under the crossing map  \eqref{crossing-saddle}. 
Cartoons of overlaps of the s-channel and t-channel unitary domains are given in figure \ref{overlaps}. They serve to set terminology for possible types of overlaps. Understanding these types helps us understand the nature of the solution space. It is also important for the analysis in section \ref{generalization}.  
For actual overlap diagrams\footnote{It turns out to be convenient to analyze this problem in $(k_{s-}, k_{s+})$ space instead of $(\Dd_s^*,\Dl_s^*)$ space. That is what we have done in appendix \ref{solving-crossing}.}, for various values of $(\za,\zb)$ and $\Dd_\phi$, see appendix \ref{solving-crossing}. 

\begin{figure}[h]
    \centering
    \subfigure[$\Dd_{\phi}=0.60$]{\includegraphics[width=0.31\textwidth]{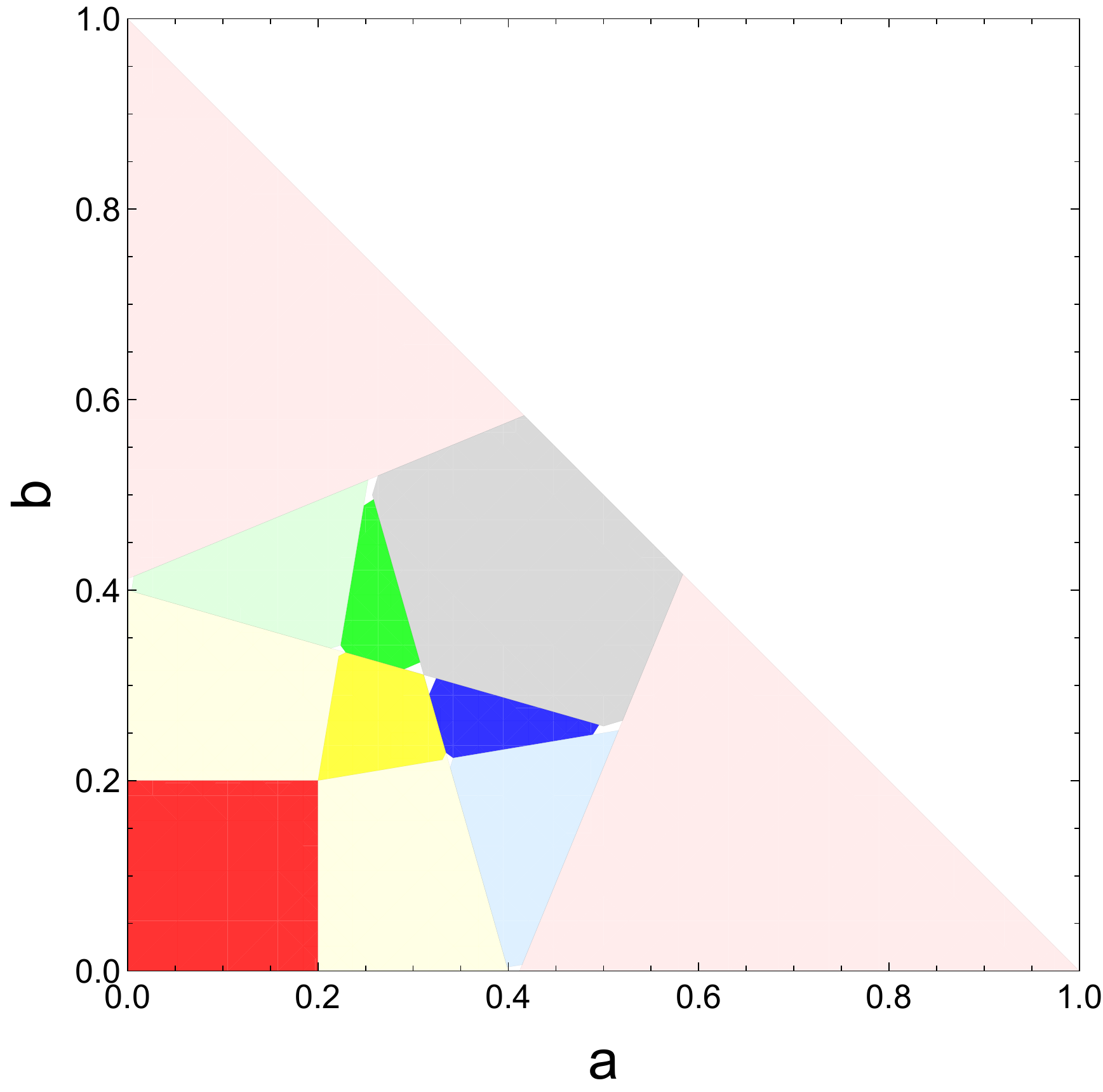} } \hspace{.1cm}
    \subfigure[$\Dd_{\phi}=0.65$]{\includegraphics[width=0.31\textwidth]{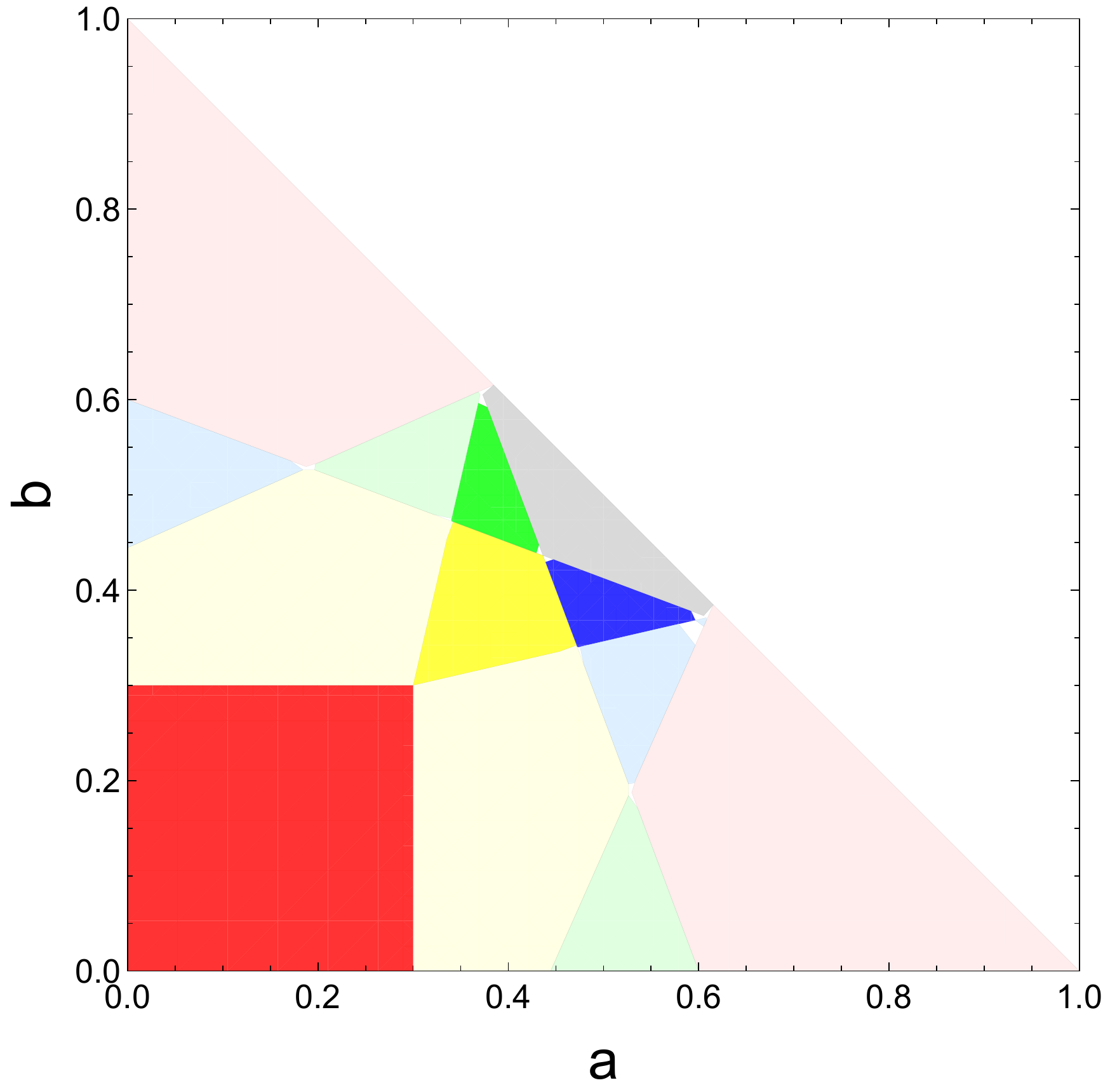} } \hspace{.1cm}
    \subfigure[$\Dd_{\phi}=0.71$]{\includegraphics[width=0.31\textwidth]{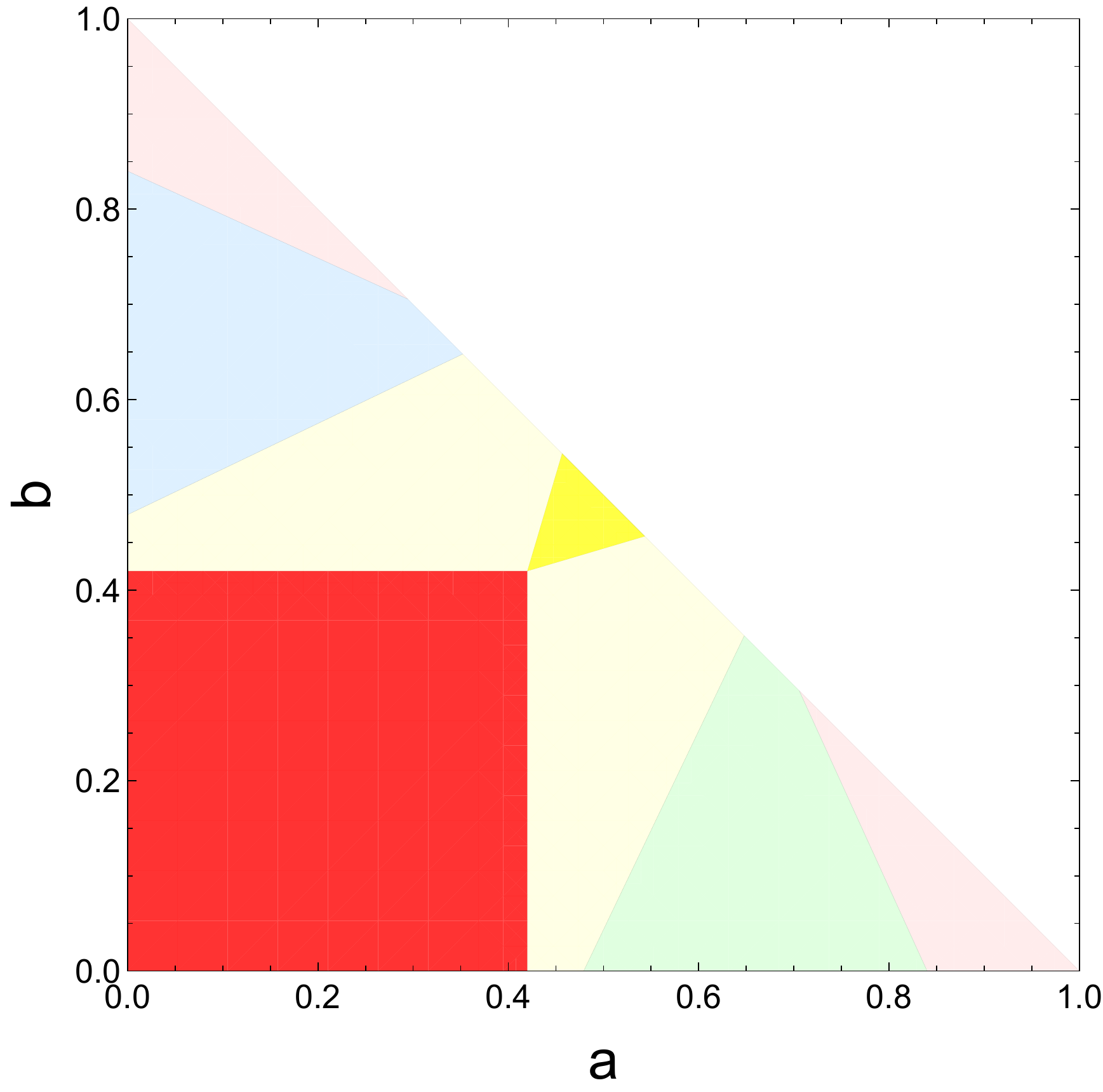}  }\hspace{.1cm}
    \subfigure[$\Dd_{\phi}=0.75$]{\includegraphics[width=0.31\textwidth]{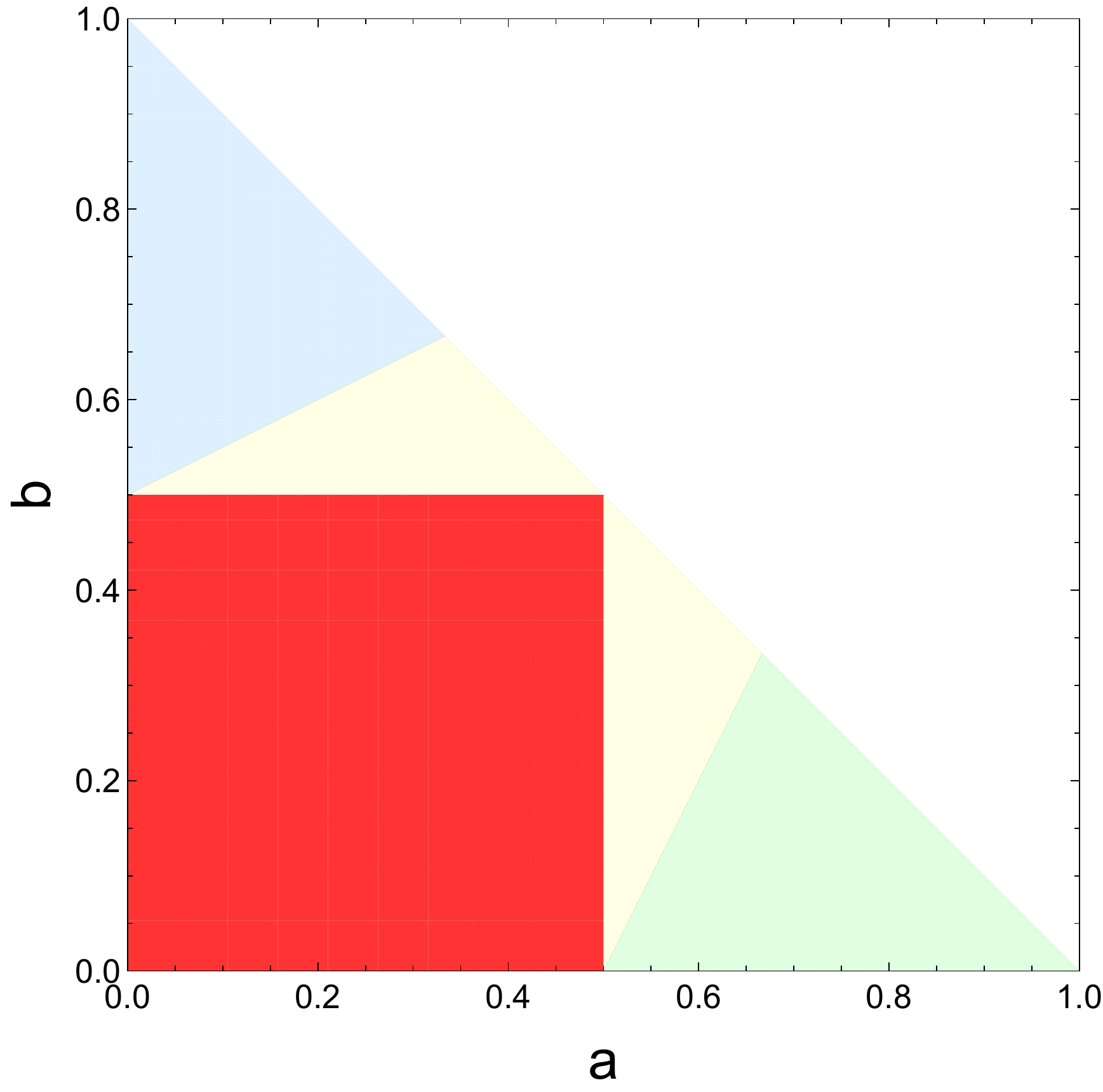}  }\hspace{.1cm}
    \subfigure[$\Dd_{\phi}=0.82$]{\includegraphics[width=0.31\textwidth]{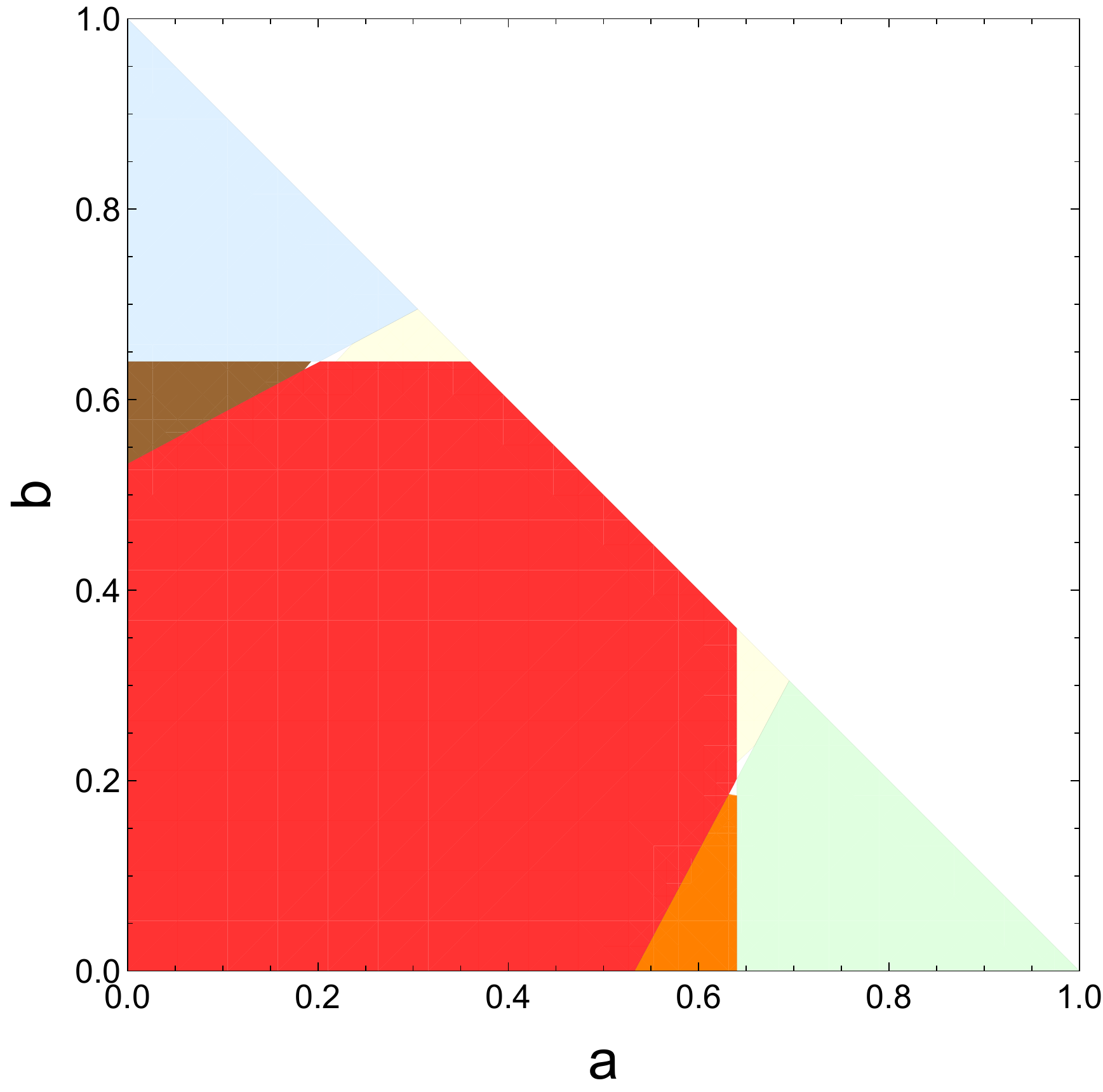}  }\hspace{.1cm}
    \subfigure[$\Dd_{\phi}=0.95$]
{\includegraphics[width=0.31    \textwidth]{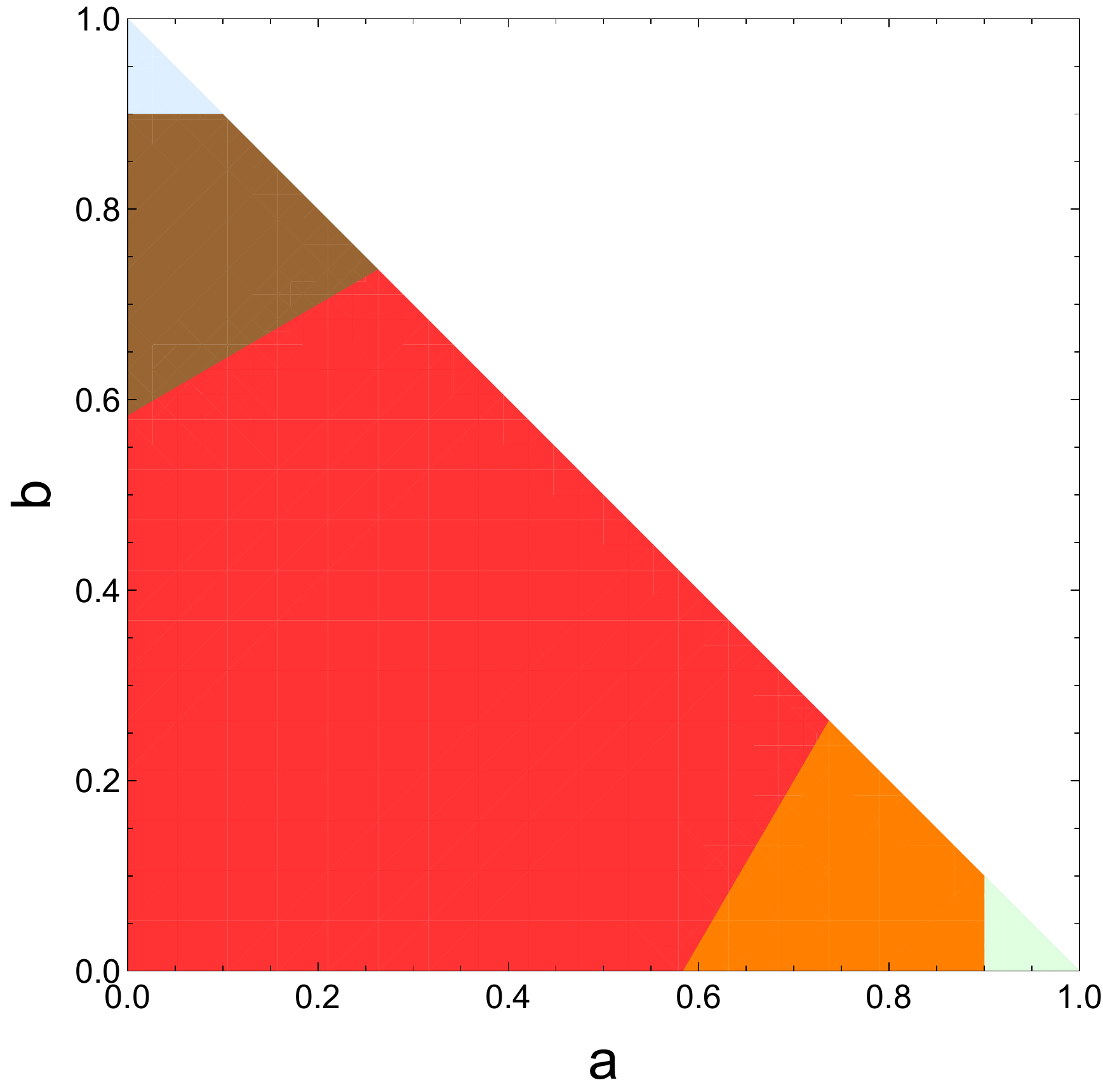}  }

  \caption{The plots show regions of $(\za , \zb)$ for different values of $\Dd_{\phi}$. The color coding is as follows. Light-pink: no solution, gray: type I, light-green: type II, light-blue: type III, green: type I + II, blue:type I + III, light-yellow: type II + III, yellow: type I + II + III, orange: type II + IV, brown: type III + IV, red: type II + III + IV.  Note that the light-pink region is present for $\Dd_\phi<3/4$. 
  }
  \label{plots}
\end{figure}

Now that we have defined the types of overlaps that can occur, we are ready to summarize the result. In figure \ref{plots}, we have colored the regions in the diamond with the kind of overlaps that are allowed by unitary and crossing as a function of $\Dd_\phi$.
Remarkably we find that for $\Dd_\phi<3/4$, there are regions of the diamond, however small,  where the crossing symmetry constraint \eqref{crossing-saddle} does not have a solution in the unitary domain. 
Lack of unitary solution to the crossing equation \eqref{crossing-saddle} means that, to leading order in large $D$, the correlator of identical scalar fields with $\Dd_\phi<3/4$ is identical to that of the GFFT! 

It turns out that this argument can be extended to make this result applicable to a wider range of external conformal dimension, $\Dd_\phi<1$ with some reasonable assumptions about the OPE density. This is as follows. 

\subsection{Extension to $\Dd_\phi <1$}\label{generalization}
In what follows we will assume that $\log(C^\epsilon_{\Dd,\Dl})$ in non-differentiable  at $(\Dd,\Dl)=(1,0)$ i.e. at the point where $\DD_1$ connects with $\DD_2$. This includes the case where there are  isolated operators i.e. operators spaced by $\CO(1)$ from the point $\Dd=1$. This assumption can be thought of as an assumption of sparseness for low lying operators.

For $3/4\leq \Dd_\phi<1$, there is no light-pink region in the diamond where any type of overlap doesn't exist. However, we note that the diamond does have regions where type IV overlap doesn't exist. These regions are light-green and light-blue regions in figure \ref{plots}. In light-green region the only overlap is of type II i.e. $\DD_2^s \cap \DD_1^t$ while in the light-blue region the overlap is of type III i.e.  $\DD_1^s \cap \DD_2^t$. If a solution to crossing equation \eqref{crossing-saddle} were to exist for $3/4\leq \Dd_\phi<1$ then it must go from being of type II in the light-green region to being of type III in the light-blue region. 

Let us first consider the light-green region. The locally dominant point in the t-channel $(\Dd_t^*,\Dl_t^*)$ must lie in $\DD_1^t$. Due to the discontinuities in $\log(C^\epsilon_{\Dd,\Dl})$, as we move in the cross-ratio space, this point moves but at most to the boundary of the smooth region and as this smooth region lies inside $\DD_1$, the locally dominant point $(\Dd_t^*,\Dl_t^*)$ continues to lie inside $\DD_1^t$. This is true even as we transit to the light-blue region. There the s-channel locally dominant point $(\Dd_s^*,\Dl_s^*)$ must necessarily be in the $\DD_1^s$. As $(\Dd_t^*,\Dl_t^*)$  lies inside $\DD_1^t$ as well, the intersection that we are looking for is of type I. But the type I solution does not exist in the light-blue region. 
  This means even for $3/4< \Dd_\phi <1$, we do not find a pair of locally dominant points that is consistent with crossing and unitarity.

\subsection{The universal saddle}\label{mft}
\begin{figure}[h]
    \centering
    \includegraphics[width=0.45\textwidth]{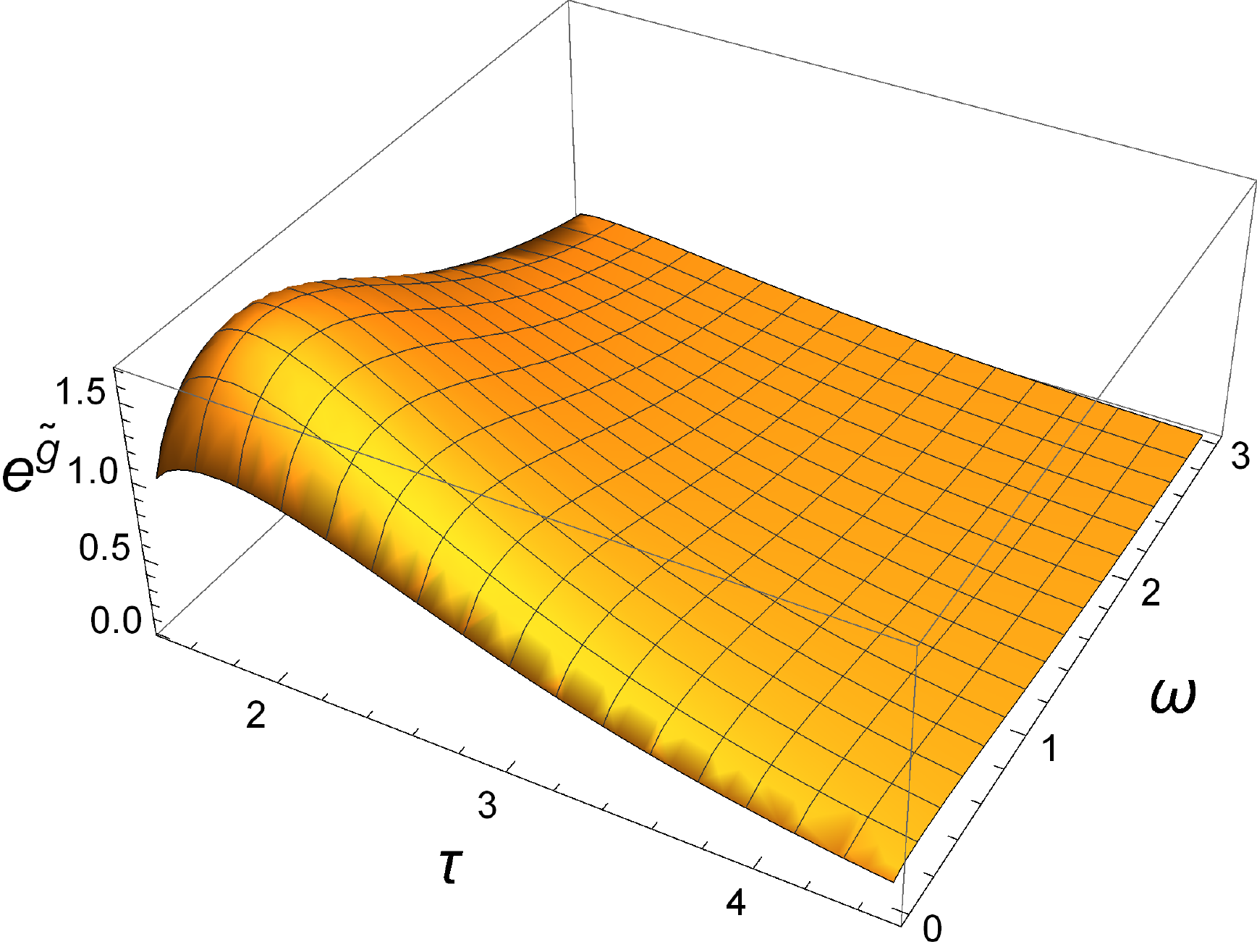} 
    \caption{Plot of $e^{\tilde g}$ in the $(\tau\equiv \Dd-\Dl, \Dl)$ plane.}
  \label{gMFT}
\end{figure}

Now that we have ruled out all possible pairs s-channel and t-channel locally dominant points that live in their respective unitary domains for $\Dd_\phi<1$, let us turn our attention to the unique  physical saddle, say in s-channel,  that \emph{can} exist. This is  dual to the contribution of the identity operator in the t-channel. The dual of the t-channel identity contribution is $(u/v)^{\Delta_\phi}$. This must come from the integration over OPE density in the s-channel.
\be
\Big(\frac{\za}{\zb}\Big)^{2\Dd_\phi D} e^{(\Du-\Dv)\Dd_\phi}=\int_{\DD_1 \cup \DD_2} d\Dd d\Dl   \Big(C_{\Dd,\Dl} \CN_{\Dd,\Dl}(\za^2,\zb^2)\Big)\,\, \CB_{\Dd,\Dl;\za,\zb}(\Du,\Dv). 
\ee
Here we have focused only on the $\ct_1$ terms as that gives the leading contribution. As mentioned earlier, the $\ct_2$ term will give its u-symmetric image namely $\za^{2 \Dd_{\phi} D}  e^{\Du \Dd_{\phi} }$. Let us assume that the integral on the right hand side is dominated by the saddle $(\Dd_s^*,\Dl_s^*)$. Matching the coefficients of $(\Du,\Dv)$ in the exponent yields, 
\be\label{univ-saddle}
(\Dd_s^*,\Dl_s^*)=(\frac12+\frac{1}{2\zb}\sqrt{(2\Dd_\phi(1+\zb)-\zb)^2-4\Dd_\phi^2 \za^2}, -\frac12+\frac{1}{2\zb}\sqrt{(2\Dd_\phi(1-\zb)+\zb)^2-4\Dd_\phi^2 \za^2}).
\ee
This could also be obtained directly by using the crossing equation \eqref{crossing-saddle} with $(\Dd_t^*,\Dl_t^*)=(0,0)$ as expected.
It is not difficult to see that this saddle belongs to the unitary region for $\Dd_\phi>1/2$ when $(\za, \zb)$ are in the diamond i.e. $\za,\zb>0, \za+\zb<1$. Matching the two sides after doing the saddle point integral and after setting $(\Du,\Dv)=(0,0)$ gives,
\be
\Big(\frac{\za}{\zb}\Big)^{2\Dd_\phi D}=\Big({\tilde f}_{\Dd_s^*,\Dl_s^*} e^{D\,{\tilde g}_{\Dd_s^*,\Dl_s^*}} \CN_{\Dd_s^*,\Dl_s^*}(\za^2,\zb^2)\Big)\sqrt{\frac{2\pi}{-D {\rm det} (g'')}}.
\ee
Here we have made manifest the $e^{D\ldots}$ dependence of the OPE density by taking  $ C_{\Dd,\Dl}\equiv {\tilde f}_{\Dd,\Dl} e^{D\,{\tilde g}_{\Dd,\Dl}}$. The notation ${\rm det} (g'')$ is a shorthand for the Hessian (the determinant of the matrix of second order partial derivatives) of $(g_{\Dd}+g_{\Dl}+{\tilde g}_{\Dd,\Dl})$ at $({\Dd_s^*,\Dl_s^*})$.
The quantity multiplying $D$ in the exponent in the conformal block is $g_{\Dd}+g_{\Dl}$. The functions $g_\Dd$ and $g_{\Dl}$ are given explicitly in appendix \ref{CB}. Inverting the relation \eqref{univ-saddle} to express $(\za,\zb)$ in terms of $(\Dd_s^*,\Dl_s^*)$ and matching the function in the exponent multiplying $D$, we get the exponential dependence in the OPE density.
\be\label{g-MFT}
{\tilde g}_{\Dd,\Dl}= \log\Big(\frac{4^{2\Dd_\phi-\Dd}\st(\Dd)\st(\Dl+\frac12)\st(\frac{\Dd-\Dl+2\Dd_\phi-2}{2})\st(\frac{\Dd+\Dl+2\Dd_\phi-1}{2})}{\st(\Dl)\st(2\Dd_\phi)\st(2\Dd_\phi-1)\st(\Dd-\frac12)\st(\frac{\Dd+\Dl-2\Dd_\phi+1}{2})\st(\frac{\Dd-\Dl-2\Dd_\phi}{2})}\Big),\qquad \quad s(x)=x^x
\ee 
This is an analytic function of $(\Dd, \Dl)$ in the unitary domain. We have plotted it in figure \ref{gMFT}.

Thus the knowledge of the saddle point as a function of $(\za, \zb)$ has allowed us to fix the OPE density completely at large $D$. After determining ${\tilde g}_{\Dd,\Dl}$, we can match the $\CO(1)$ function that multiplies $e^{D\ldots}$ terms on both sides. This allows us to compute ${\tilde f}_{\Dd,\Dl}$.  Simplifying this function to a closed-form turns out to be difficult. We have computed it numerically.

As the universal saddle point is dual to identity operator, we expect the OPE density ${\tilde f}e^{D{\tilde g}}$ obtained above to be the large $D$ limit of the OPE density of GFFT. The GFFT OPE coefficients are known in closed form in arbitrary dimensions \cite{Fitzpatrick:2011dm}. We reproduce them below for reference. 
\bea
\tCC_{\Delta,\ell}&=&\frac{[1+(-1)^\ell](\Delta_\phi-D/2+1)^2_n(\Delta_\phi)^2_{\ell+n}}{\ell!n!(\ell+D/2)_n(2\Delta_\phi+n-D+1)_n(2\Delta_\phi+2n+\ell-1)_\ell(2\Delta_\phi+n+\ell-D/2)_n}\nonumber\\
&=& \frac{[1+(-1)^{\omega D}] \left[ \left( \left(\delta_{\phi}-\frac{1}{2} \right) D+1 \right)_{\alpha D}\right]^{2} [(\delta_{\phi}D)_{(\alpha + \omega)D}]^{2} }{ \left( \left(\omega + \frac{1}{2} \right)D \right)_{\alpha D} \left( \left( 2 \delta_{\phi} +\alpha -1 \right)D +1 \right)_{\alpha D}  }\nonumber\\
&\times &\frac{(\Gamma[\omega D +1] \Gamma[\alpha D +1])^{-1}}{\left( \left( 2 \delta_{\phi} + 2 \alpha + \omega \right)D -1 \right)_{\omega D} \left( \left( 2 \delta_{\phi} + \alpha + \omega -\frac{1}{2} \right)D \right)_{\alpha D}}.
\eea
In the second line we have separated all the $D$ dependence used to take the scaling limit. 
The large $D$ limit is taken using Stirling's approximation for the Gamma function,
\be
\Gamma(z) \xrightarrow{z\to \infty}\sqrt{\frac{2 \pi}{z}}\Big(\frac{z}{e}\Big)^{z}\Big(1+\CO(\frac1z)\Big).
\ee
In the large $D$ limit the GFFT OPE density takes the form ${\tilde f}_{GFFT} e^{D {\tilde g}_{GFFT}}$. The function ${\tilde g}_{GFFT}$ is precisely ${\tilde g}_{\Dd,\Dl}$ given in \eqref{g-MFT}. Also,
\be\label{f-MFT}
{\tilde f}_{\Dd,\Dl}= \frac{1}{\pi D }\Big(\frac{\Dd_\phi^2 (2\Dd-1)(\Dd+\Dl)(\Dd-\Dl-1)(\Dd+2\Dd_\phi-\Dl-2)(\Dd-2\Dd_\phi+\Dl+1)}{(2\Dd_\phi-1)^2\Dd^3\Dl(\Dd-2\Dd_\phi-\Dl)(\Dd+2\Dd_\phi+\Dl-1)(1+2\Dl)}\Big)^\frac12,
\ee
and it agrees with ${\tilde f}_{\Dd,\Dl}$ computed numerically using crossing symmetry.

The u-symmetric contribution $u^{\Delta_\phi}$ is subleading in the diamond and comes from a saddle point that lies in the complex $(\Dd,\Dl)$ space. One can indeed check by taking the  OPE density \eqref{g-MFT}, \eqref{f-MFT} that the $\ct_2$ saddle precisely gives this contribution.

\section{Discussion and outlook}\label{discussion}

In this paper, we have argued that, at large $D$,  the four-point function of identical scalar operators with $\Dd_\phi<1$ is the same as that in the GFFT to leading order. 
First, we note that it is unreasonable to expect this result to extend beyond $\Dd_\phi<1$. This is for the following reason. Consider a tensor product of GFFT of $N$ fields $\chi_i$ such that they have the same conformal dimension $\Delta_\chi=\Delta_\phi/2$. 
This theory has $SO(N)$ global symmetry. Importantly, as $\Dd_\phi>1$, this GFFT is unitary. 
Consider the four-point function of flavor singlet operator $\chi_i\chi_i$. We define this to be the operator $\phi$ whose four-point function we consider. By construction, the conformal dimension of $\phi$ is $\Delta_\phi$. This four-point function of the composite operator is computed by all the Wick contractions. The stripped four-point function is
\be
G(u,v)=1+\Big(\frac{u}{v}\Big)^{\Delta_\phi}+u^{\Delta_\phi}+ \frac{1}{N}\Big(u^{\Delta_\phi/2}+\Big(\frac{u}{v}\Big)^{\Delta_\phi/2}+\frac{u^{\Delta_\phi}}{v^{\Delta_\phi/2}}\Big).
\ee
Here we have normalized the two-point function of $\phi$ to be $1$. First three terms come from the disconnected diagrams. This is the same as what would appear in the GFFT of $\phi$ itself. The next three terms come from connected diagrams and can have a relative factor compared to the disconnected piece.

At large $D$, the disconnected piece comes from the identity operator and the two saddle point that are discussed in section \ref{mft} while the connected terms come from three new saddle points. Among the three terms, $(u/v)^{\Delta_\phi/2}$ is leading in the diamond. Interestingly, it is self-dual under crossing symmetry. Substituting $(u,v)=(\za^2 e^{\Du/D}, \zb^2 e^{\Dv/D})$ and matching coefficients of $\Du$ and $\Dv$ in the exponent with the form \eqref{largeDblock}, we see that the saddle point is at $(k_{s-},k_{s+})=(k_{t-},k_{t+})=(\Dd_\phi/4,\Dd_\phi/4)$. In terms of $(\Dd, \Dl)$ this means,
\bea\label{connected}
(\Dd_t^*,\Dl_t^*)&=&(\Dd_s^*,\Dl_s^*)\nonumber\\
(\Dd_s^*,\Dl_s^*)&=&\Big(\frac12+\frac{1}{2\zb}\sqrt{(\Dd_\phi(1+\zb)-\zb)^2-\Dd_\phi^2 \za^2}, -\frac12+\frac{1}{2\zb}\sqrt{(\Dd_\phi(1-\zb)+\zb)^2-\Dd_\phi^2 \za^2}\Big).
\eea
Interestingly this saddle point lies in the unitary domain for $\Dd_\phi\geq 1$ and when $(\za,\zb)$ are in the diamond i.e. for $\za,\zb>0, \za+\zb<1$. This is consistent with our analysis because precisely for $\Dd_\phi>1$, crossing and unitarity  allow other solutions apart from the universal one and \eqref{connected} is one of them.

However, one may wonder whether this is the only other solutions to the crossing equation \eqref{crossing-saddle} for $\Dd_\phi> 1$ (and $\Dd_\phi<3/2$)\footnote{For $\Dd_\phi>3/2$, in addition to the unitary GFFT of $\chi$ such that $\chi^2\equiv \phi$,  one could consider unitary GFFT of yet another operator $\mu$ such that $\mu^3\equiv \phi$ and so on.}. This would mean that for $\Dd_\phi>1$, the GFFT of $\phi$ and GFFT of $\chi$ (where $\phi$ is the composite operator $\chi^2$) are the only two solutions at large $D$. This is an interesting possibility but one about which we can't say anything currently. Of course, it is entirely possible that for $\Dd_\phi>1$, things may not be as simple as that. Note that we are asking this question without imposing the existence of stress tensor. Having a stress tensor could provide additional constraints, more severely constraining the solutions at large $D$.

In this paper, we have constrained the unitary solutions to the crossing for $\Dd_\phi <1$ only to leading order at large $D$. At sub-leading order, the solution may receive perturbative $1/D$ corrections in addition to the non-perturbative ones. The non-perturbative corrections correspond to subleading saddle points. As discussed below equation \eqref{largeDcorrelator}, we already know of a mechanism by which these could appear, namely, from the rapidly oscillating part of the OPE density (which we have chosen to smear over). We believe that these corrections would be very difficult to control. However, the perturbative corrections can be accounted for relatively straightforwardly. They come from $1/D$ corrections to the conformal blocks as well as $1/D$ corrections to the smeared OPE density. The corrections to the conformal blocks can be computed from the conformal Casimir equation. The question of computing $1/D$ corrections to the correlator then is actually the question of controlling the $1/D$ corrections to the smeared OPE density.  
As the contribution of the saddle point that is dual to the identity operator is completely fixed to all orders in $1/D$, the $1/D$ corrections to the smeared OPE coefficient density must also be fixed. It would be interesting to compute these corrections explicitly and match them with the GFFT OPE coefficients in $1/D$ expansion. 

Our arguments in the paper have uniquely fixed the leading order correlator only in the Lorentzian diamond. Elsewhere in the cross-ratio space, the conformal blocks could rapidly oscillate in phase and hence the dominating saddle could be anywhere in the complex $(\Dd,\Dl)$ space. That is why it would seem difficult to extend the results outside the diamond.  However, 
in addition to fixing the correlator in the diamond, we are also able to fix the leading order OPE density (see section \ref{mft}). 
This strongly suggests that the theory of operator $\phi$ is perhaps GFFT itself which would make $\phi$ decoupled from the rest of the theory. Phrased another way, our results suggest that OPE coefficients of relevant operators in 
a non-trivial conformal theory\footnote{not necessarily having a stress tensor} in large dimensions must be exponentially suppressed at large $D$.

Even though we are considering $D$ a formal parameter in solving conformal bootstrap equations, there are indications that the unitary solution space to crossing is drastically different for non-integer $D$ compared to integer $D$, see \cite{Hogervorst:2015akt, Ji:2018yaf}. In our analysis, we have not made any assumptions about integrality of $D$ and the large $D$ limit could as well be taken with $D$ an integer. It would be useful to understand how the nature of the solution space changes as we make $D$ fractional in our approach. 

We would also like to point out the paper \cite{Haldar:2019prg} where authors show that the number of subtractions to write a dispersion relation goes to infinity as $D$ goes to infinity. This is perhaps an indication that such theories prefer to be free \cite{personal}.

In this paper, the object that played an important role is the smeared OPE density. A more rigorous analysis of the constraints on the smeared OPE density may be possible using the so-called ``Tauberian theorems". These techniques have recently been applied to conformal field theories to essentially estimate the errors associated with such a smearing \cite{Qiao:2017xif, Mukhametzhanov:2018zja, Mukhametzhanov:2019pzy, Pal:2019zzr}. It would be nice to put our treatment on a more solid footing with the use of similar techniques. 
Finally, it would be interesting to approach the problem of CFTs in large dimensions in other ways, for example using the Lorentzian inversion formula \cite{Caron-Huot:2017vep} or using the method of extremal functionals which has provided optimal analytic bounds on OPE coefficients in the large $\Delta$ limit \cite{Mazac:2018mdx}.

\section*{Acknowledgements}
We would like to thank Thomas Dumitrescu, Subham Duttachowdhury,  Indranil Halder, Ashoke Sen,  David Simmons-Duffin, Aninda Sinha and Balt van Rees for useful discussions. 
We would also like to thank Subham Duttachowdhury for comments on the manuscript. We are especially thankful to Shiraz Minwalla for inspiring discussions and also for useful comments on the manuscript.
The work of both the authors was supported by the Infosys Endowment for the study of the Quantum Structure of Spacetime.
The work of A.G. is also supported by the SERB Ramanujan fellowship. A.G. would like to acknowledge that part of this work was performed at the Aspen Center for Physics, which is supported by the National Science Foundation grant PHY-1607611.  We would all also like to acknowledge our debt to the people of India for their steady support to the study of the basic sciences.

\appendix

\section{Explicit blocks in large $D$}\label{CB}

The conformal blocks have been computed in the large $D$ limit as
\bea
\CF_{\Delta,l}(u,v)&=&\frac{2^{\Delta+l}}{\sqrt{y_--y_+}}A_{\Delta}(y_+)A_{1-l}(y_-),\qquad y_{\pm}=\frac{u}{(1\pm \sqrt{v})^{2}}\nonumber\\
A_x(y)&=&y^{\frac{x}{2}} \,_2F_1(\frac{x-1}{2},\frac{x}{2},x-\frac{D}{2}+1;y)
\eea

We want to scale the conformal dimension and spin with $D$ as $\Delta\rightarrow \Dd D$ and $l \rightarrow \Dl D$, as explained in the main text. 

The conformal dimension part of the block, $A_{\Delta}(y_+)$ is estimated using the Euler integral representation of the conformal block.
\be
_2F_1(a,b,c;z)=\frac{\Gamma(c)}{\Gamma(b)\Gamma(c-b)}\int_0^1 dt\, t^{b-1}(1-t)^{c-b-1}(1-zt)^{-a}
\ee
In our large $D$ limit and for Hypergeometric function in $ A_\Delta(y_+)$ the integral reduces to a saddle point integral.
The saddle point is at
\be
t_*= t^{\pm}_{\Dd}=\frac{2 \Dd -1 \pm \sqrt{4 (\Dd -1) \Dd  (1-y_+)+1}}{2 (\Dd -1) y_+}
\ee
Now, $t_*= t^{-}_{\Dd}$ is chosen because it lies between $(0,1)$ when $\Dd > 1/2$. Thus, at leading order in $1/D$,
\bea
\CF_\Dd \equiv  2^{\Delta} A_\Delta(\frac14)&=&\frac{ \Gamma \left[D(\Dd-\frac12) +1 \right]}{\Gamma \left[ \frac{\Dd}{2} D \right] \Gamma \left[ (\frac{\Dd-1}{2}) D +1 \right]} \frac{ \sqrt{1- y_{+} t^{-}_{\Dd}} }{t^{-}_{\Dd}} \Big( 2^\Dd y_{+}^{\frac{\Dd}{2}} (t^{-}_{\Dd})^{\frac{\Dd}{2}}(1-t^{-}_{\Dd})^{\frac{\Dd-1}{2}}(1-y_+ t^{-}_{\Dd})^{-\frac{\Dd}{2}}\Big)^{D}\sqrt{\frac{2 \pi}{-D \beta}}\nonumber\\
&=& \nonumber\\
{\rm where}\quad \beta&=& \frac{\partial^2}{\partial t^2} \log[t^{\frac{\Dd}{2}}(1-t)^{\frac{\Dd-1}{2}}(1-y_+ t)^{-\frac{\Dd}{2}}] |_{t=t^{-}_{\Dd}}
\eea \\
Putting $(u,v)=(\za^2 e^{\frac{\Du}{D}},\zb^2 e^{\frac{\Dv}{D}})$ implies $y_+ = \frac{\za^2}{(1+\zb)^2} (1+ \frac{1}{D}
\frac{\zb}{1+\zb}( \frac{(1+\zb)}{\zb}\Du-\Dv) + \CO[1/D]) $. Using Sterling approximation for $\Gamma[x] \stackrel{x\to \infty}{=}   \sqrt{\frac{2 \pi}{x}} \  x^{x} e^{-x} \ [1 + \CO (\frac{1}{x})] $, the conformal dimension dependent part of block $\CF_{\Dd}$ at leading order in $1/D$ reduces to,

\begin{equation}
\CF_{\Dd}= f_{\Dd} \ e^{D g_{\Dd}} e^{k_{+}(\za,\zb,\Dd) ((1+\frac{1}{b}) \Du - \Dv)}
\end{equation}
where,
\begin{align*}
&g_{\Dd}= \log \Big[ \sqrt{\frac{2(\Dd -1)^2 \hat{y}_{+}}{\left(2 \Dd -1\right) (A-2 \Dd +2 (\Dd -1) \hat{y}_{+}+1)}} \left(\frac{2 (2 \Dd -1)^2 (A-2 \Dd +1) (A-2 \Dd +2 (\Dd -1) \hat{y}_{+}+1)}{(1-A)\Dd (\Dd -1)^2   \hat{y}_{+}}\right)^{\Dd /2} \Big] \\
&f_{\Dd}= \sqrt{\frac{(\Dd -1)^{-3}(A-1)^3 \Dd  (2 \Dd -1) (A-2 \Dd +2 (\Dd -1) \hat{y}_{+}+1)^2}{2  \left(16 \Dd  (\delta -1)^2 (A-\Dd ) \hat{y}_{+}^2 +16 \Dd  (\Dd -1)  (A-\Dd ) (A-2 \Dd +1) \hat{y}_{+} + (A-2 \Dd +1)^2 \left(4 A \Dd +(A-1)^2-4 \Dd ^2\right) \right)}}
\end{align*}
with, $ A=\sqrt{1+4 (1-\hat{y}_{+}) \Dd (\Dd -1)} $ and $ \hat{y}_{+} = \frac{\za^2}{(1+\zb)^2} $.

Similar steps can be done to find the spin dependent part of the block at $y_{\pm} = \frac{\za^2 \ e^{\frac{\Du}{D}}}{(1 \pm \zb \  e^{\frac{\Dv}{2 D}})^2}$ in leading order in $1/D$ as:
\begin{equation}
\CF_\Dl = \frac{2^{l}}{\sqrt{y_- - y_+}} A_{1-l}(y_-) = f_{\Dl} \ e^{D g_{\Dl}} e^{k_{-}(\za,-\zb,1+\Dl) ((1-\frac{1}{\zb}) \Du - \Dv)}
\end{equation} 
where,
\begin{align*}
&g_{\Dl} = \log \Big[ \sqrt{\frac{2(\Dl +1)^2 \hat{y}_{-}}{\left(2 \Dl +1\right) (B- 2 \Dl +2 (\Dl +1) \hat{y}_{-}-1)}} \left(\frac{ (2 \Dl 
+1)^2 (B-2 \Dl -1) (B-2 \Dl +2 (\Dl +1) \hat{y}_{-}-1)}{- 8\Dl (1+B) (\Dl +1)^2   \hat{y}_{-}}\right)^{-\Dl /2} \Big] \\
&f_{\Dl} =  \sqrt{\frac{- (4 \hat{y}_{-})^{-1} (\hat{y}_{-} - \hat{y}_{+})^{-1}(\omega +1)^{-5} (B+1)^2 (2 \omega +1)^3 (B-2 \omega -1) (B-2 \omega +2 (\omega +1) \hat{y}_{-}-1)^3}{ 16 \omega  (\omega +1)^2  (B-\omega )\hat{y}_{-}^2 +16 \omega  (\omega +1)  (B-\omega ) (B-2 \omega -1)\hat{y}_{-} +(B-2 \omega -1)^2 \left(4 B \omega +(B+1)^2-4 \omega ^2\right) }}
\end{align*}
with, $ B=\sqrt{1+4 (1-\hat{y}_{-}) \Dl (\Dl + 1)} $ and $ \hat{y}_{\pm} = \frac{\za^2}{(1 \pm \zb)^2} $.

Thus, the overall conformal block is: $\CF_{D \Dd, D \Dl}=\CF_{\Dd} \ \CF_{ \Dl}$ .

\subsection{Scalar block}

\be\label{scalar-series}
\CF_{\Delta,0}=\sum_{m,n=0}^{\infty}\frac{(\Delta/2)_n^2(\Delta/2)_{n+m}^2}{(\Delta+1-D/2)_n(\Delta)_{2n+m}}\frac{u^{\frac{\Delta}{2}+n}}{n!}\frac{(1-v)^{m}}{m!}, \qquad (x)_n\equiv \Gamma(x+n)/\Gamma(x).
\ee
We will scale the conformal dimension $\Delta$, $n$ and $m$ with $D$ as $\Delta\rightarrow \Dd D$, $n \rightarrow \alpha D$ and $m \rightarrow \beta D$ respectively. With this scaling the sum over $n,m$ becomes integral over $\alpha,\beta$ which can be performed using saddle point integration. Substituting $(u,v)=(\za^2 e^{\frac{\Du}{D}},\zb^2 e^{\frac{\Dv}{D}})$ and again using Sterling approximation for $\Gamma(x) \stackrel{x\to \infty}{=}   \sqrt{\frac{2 \pi}{x}} \  x^{x} e^{-x} \ [1 + \CO (\frac{1}{x})] $, we get the saddle point $(\alpha^*,\beta^*)$ as:
\bea
&\alpha^* = \Big( 1+\frac{1}{\zb}\Big) k_{+}(\za,\zb,\Dd) -\frac{\Dd}{2}  \\
&\beta^*  = \Big( \frac{1-\zb^2}{\zb^2} \Big) k_{+}(\za,\zb,\Dd)
\eea
which on substituting back gives the required $\Du,\Dv$ dependence as well as the normalization part of the scalar block as mentioned in the main text.

\section{Generalities of the saddle point}\label{saddle-general}

Consider the following integral in the large $D$ limit,
\be
\CI=\int_{a}^b \,f(x)\, e^{D \, g(x)}.
\ee
This is known as the integral of the Laplace type. As $D$ is large, if $g(x)$ is a real function in the integration range $[a,b]\in {\mathbb R}$, the integral is dominated by the point where $g(x)$ takes maximum value. This point could be in the interior of the integration domain or on the boundary. In both cases, integral gets dominant contribution from the neighborhood of that point. At leading order in large $D$,
\begin{itemize}
\item when $g(x)$ is maximum in the interior
\be\label{saddlept}
\CI\sim \sqrt{\frac{2\pi}{-D g''(x^*)}} \,f(x^*)\, e^{D g(x^*)},
\ee
\item when $g(x)$ is maximum outside\footnote{This formula blows up when the maximum is exactly on the boundary i.e. when either $g'(a)=0$ or $g'(b)=0$ but this is outside the validity of this formula. The formula is applicable for $g'(a),g'(b) \geq \CO(1/D)$. }
\be\label{endpt}
\CI\sim {\rm max}\Big(\frac{-1}{D\, g'(a)}\, f(a) \, e^{D g(a)}, \frac{1}{D\, g'(b)}\, f(b) \, e^{D g(b)}\Big).
\ee
\end{itemize}
The integral of the first type is called the saddle point integral. Of course, the integral could have other local maxima and the integral receives similar but subdominant contribution from those. If the function $g(x)$ is piecewise continuous, we divide the integration range in these pieces and the above discussion applies to the locally dominant point from each piece.

We want to highlight the non-analytic behavior of the integral as the dominant point transits across the integration range. To that end, consider a Laplace type integral with a parameter $\lambda$.  
\be
I(\lambda)=\int_0^\infty dx\, e^{-D (x-\lambda)^2}
\ee
For $\lambda > 0$, the integral is computed using the formula \eqref{saddlept}. This gives $I(\lambda)=\sqrt{\pi}$. 
While for $\lambda <0$, the integral is computed using the formula \eqref{endpt}. This gives $I(\lambda)=e^{-D\lambda^2}/(2D|\lambda|)$. It is clear that $I(\lambda)$ is non-analytic at  $\lambda=0$ i.e. at the point where the saddle point enters the integration range. In fact the integral increases in magnitude as the maximum transitions inside the integration range.

On the other hand, if $g(x)$ has a varying imaginary part in the integration range then the integrand has a rapidly oscillating phase. Naturally such an integral is difficult to estimate by staying on the real line as the oscillating phase is expected to cancel huge numbers to give tiny remainders. The way out is to use the analyticity of the integrand to deform the contour into the complex plane such that the new contour of integration is on the path of stationary phase i.e. the path of constant ${\rm Im}[g(z)]$. This is also the path of steepest descent for ${\rm Re}[g(z)]$. Here we want the reader to note that as a result of contour deformation, the saddle point could be outside the integration range, even in the complex plane.

\section{Solving the crossing equation}\label{solving-crossing}

As discussed in the main text unitarity and crossing implies both the saddles $(\Dd_{s}^{*},\Dl_{s}^{*})$ and $(\Dd_{t}^{*},\Dl_{t}^{*})$, related to each other by equation \eqref{crossing-saddle}, must exist in $\DD_1 \cup \DD_2$. Let us see what these constraints mean for $(k_{s-}, k_{s+})$. Inverting the map \eqref{largeDblock}  $(\Dd^{*},\Dl^{*})\to (k_{-}, k_{+})$, we get
\bea
(\Dd^{*},\Dl^{*})&=&(\zeta(k_{+},\za,\zb), - \zeta(k_{-},\za,-\zb))\\
{\rm where}\quad \zeta(x,\za,\zb) &=& \frac{1}{2} + \frac{1}{2 \zb} \sqrt{ (4x(1+b)-b)^2-(4x a)^2 }.\nonumber
\eea
Using this equation as well as \eqref{crossing-saddle}, we map the regions $\DD_1^s, \DD_2^s, \DD_1^t, \DD_2^t$ in $(k_{s+},k_{s-})$ space. 
\begin{align}\label{conditions}
\nonumber
& \DD_1^s = \{ k_{s-}=\CQ_{\Dl_s} ( k_{s+}, \za ,\zb ) ,  \tilde{\CQ}_{\Dd_s} ( k_{s-}, \za ,\zb ) \leq k_{s+} < \CQ_{\Dd_s} ( k_{s-}, \za ,\zb ) \} \\
\nonumber 
& \DD_2^s = \{ k_{s-}>\CQ_{\Dl_s} ( k_{s+}, \za ,\zb ) ,  k_{s+} \geq \CQ_{\Dd_s} ( k_{s-}, \za ,\zb ) \} \\
\nonumber 
&\DD_1^t = \{ k_{s-}=\CQ_{\Dl_t} ( k_{s+}, \za ,\zb ) ,  \tilde{\CQ}_{\Dd_t} ( k_{s-}, \za ,\zb ) \geq k_{s+} > \CQ_{\Dd_t} ( k_{s-}, \za ,\zb ) \} \\ 
& \DD_2^t = \{ k_{s-}>\CQ_{\Dl_t} ( k_{s+}, \za ,\zb ) ,  k_{s+} \leq \CQ_{\Dd_t} ( k_{s-}, \za ,\zb ) \} \ , 
\end{align}
The $\CQ$ functions appearing above are,
\begin{align}
\nonumber
&\CQ_{\Dl_s} ( k_{s+}, \za ,\zb )  = 0 \\
\nonumber
&\CQ_{\Dl_t} ( k_{s+}, \za ,\zb ) = \frac{-\zb \  \delta_\phi + (1+\za+\zb) k_{s+}}{(1+\za-\zb)} \\
\nonumber
&\CQ_{\Dd_s} ( k_{s-}, \za ,\zb ) = \frac{\zb(1+\zb) + \sqrt{\gamma (k_{s-})}}{4 ((1 + \zb)^2 - \za^2 )} \\
\nonumber
&\CQ_{\Dd_t} ( k_{s-}, \za ,\zb )  =  \frac{(2 \Dd_{\phi} -1) \za^2-2 \Dd_{\phi}  (\zb+1)^2+\zb+1 + \sqrt{\eta_2 \ k_{s-}^2 + \eta_1 \ k_{s-} +\eta_0 } }{4 (\za-\zb-1) (\za+\zb+1)}\\
\nonumber
&\tilde{\CQ}_{\Dd_s} ( k_{s-}, \za ,\zb ) = \frac{\zb}{4 (1 - \za + \zb)}  \\
&\tilde{\CQ}_{\Dd_t} ( k_{s-}, \za ,\zb ) = \frac{2 k_{s-} \big( (1 - \zb)^2 - \za^2 \big)  + 2 \zb \big( 1 + \za - \zb \big) \delta_ \phi - \za \zb}{2 \big(1 - (\za - \zb)^2 \big)} \label{Unitarity_in_k} 
\end{align}
where,
\begin{align*}
& \gamma (k_{s-}) = \zb^2 (1 + \zb)^2 - 8 \zb (1 - \zb) \big( (1 + \zb)^2 - \za^2 \big) k_{s-} + 16 \big((\zb+1)^2-\za^2 \big) \big((1-\zb)^2-\za^2 \big)k_{s-}^2 \\
& \eta_{0} = (2\Dd_{\phi} -1 )^2 \za^4-2 (2 \Dd_{\phi} -1) \za^2 \left(2 \Dd_{\phi}  \left(\zb^2+1\right)-\zb-1\right)+\left(2 \Dd_{\phi}  \left(\zb^2-1\right)+\zb+1\right)^2 \\
& \eta_{1} = -8 \left((\za-\zb-1) (\za+\zb+1) \left((2 \Dd_{\phi} -1) \za^2-2 \Dd_{\phi}  (\zb-1)^2-\zb+1\right)\right) \\
& \eta_{2} = 16 \left(\za^4-2 \za^2 \left(\zb^2+1\right)+\left(\zb^2-1\right)^2\right)
\end{align*}

Conditions \eqref{conditions} are used to find regions in the diamond where overlaps of various types occur. These are then used to get figure \ref{plots}. We have illustrated some examples of  overlap diagrams in $(k_{s-},k_{s+})$ space and the corresponding regions in the diamond in figures \ref{type-IV}, \ref{type-III}, \ref{type-II} and \ref{type-I}. We have taken $\Dd_\phi=0.6$ in all of these plots. 
The blue and the red regions (in the part(a) of each figure) are the unitarity domains of s-channel and t-channel saddle points respectively. The dashed blue  and blue curves correspond to the s-channel constraints $k_{s-}=\CQ_{\Dl_s} ( k_{s+}, \za ,\zb )$ and $k_{s+}=\CQ_{\Dd_s} ( k_{s-}, \za ,\zb )$ respectively. The dashed red and red curves correspond to the t-channel constraints $k_{s-}=\CQ_{\Dl_t} ( k_{s+}, \za ,\zb )$ and $k_{s+}=\CQ_{\Dd_t} ( k_{s-}, \za ,\zb )$ respectively. 
The endpoint of the dashed blue and dashed red curve are marked with $\tt P$urpole and $\tt O$range points. They correspond to $  k_{s+}=\tilde{\CQ}_{\Dd_s} ( k_{s-}, \za ,\zb )$ and $k_{s+}= \tilde{\CQ}_{\Dd_t} ( k_{s-}, \za ,\zb )$ respectively.
It is convenient to label the intersections of these curves also.  
The $\tt B$lue, $\tt R$ed and $\tt G$reen points are the intersection of curves (blue $\cap$ dashed blue), (red $\cap$ dashed red) and (dashed blue $\cap$ dashed red) respectively. The coordinates of these points are
\bea
(k_{s-}^{\tt P},k_{s+}^{\tt P})&=& (0 , \frac{\zb}{4 (1 - \za + \zb)} )\nonumber\\
(k_{s-}^{\tt O},k_{s+}^{\tt O})&=& ( \frac{\Dd_{\phi}}{2} - \frac{\za + \zb + 1}{8 (1 + \za - \zb)} , \frac{4 \Dd_{\phi} -1 }{ 8 } ) \nonumber\\
(k_{s-}^{\tt B},k_{s+}^{\tt B})&=& (0 , \frac{\zb (1 + \zb)}{2 ( (1 + \zb)^2-\za^2)})\nonumber\\
(k_{s-}^{\tt R},k_{s+}^{\tt R})&=& ( \frac{\Dd_{\phi}}{2} - \frac{\za +1}{4 (1 + \za - \zb)} , \frac{\Dd_{\phi}}{2} - \frac{\za +1}{4 (1 + \za + \zb)})\nonumber\\
(k_{s-}^{\tt G},k_{s+}^{\tt G})&=& (0 , \frac{\zb \ \Dd_{\phi} }{ 1 + \za + \zb }  ) .
\eea
Part (b) of each figure shows the region of existence of corresponding type of overlap in the Lorentzian diamond (i.e. $0 \leq \za,\zb, \za+\za<1 $). This is the part below the diagonal black line.

\subsection*{Existence of type-IV overlap}

\begin{figure}[t]
    \centering
    \subfigure[]{\includegraphics[width=0.35\textwidth]{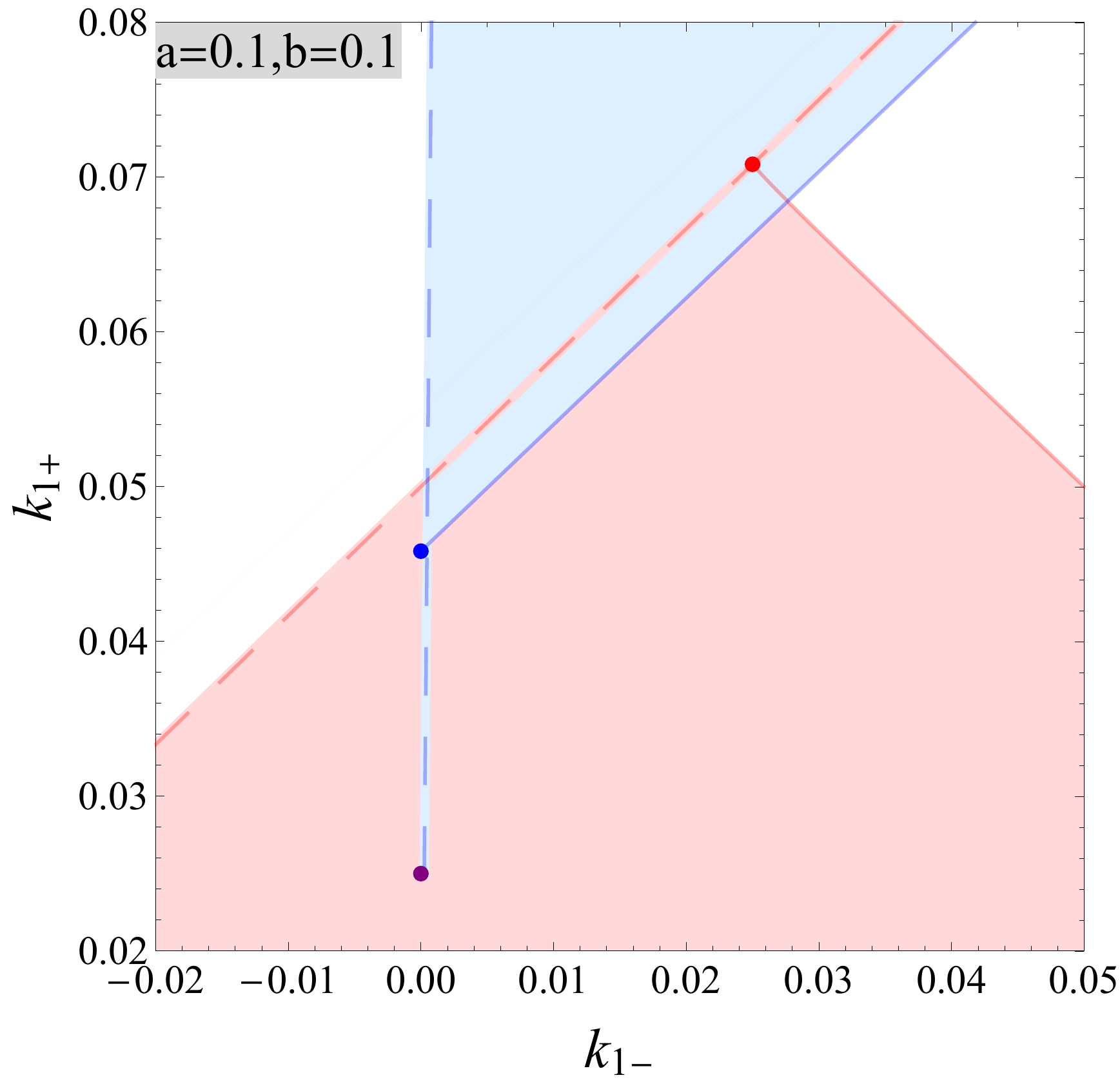} } \hspace{.5cm}
    \subfigure[]{\includegraphics[width=0.35\textwidth]{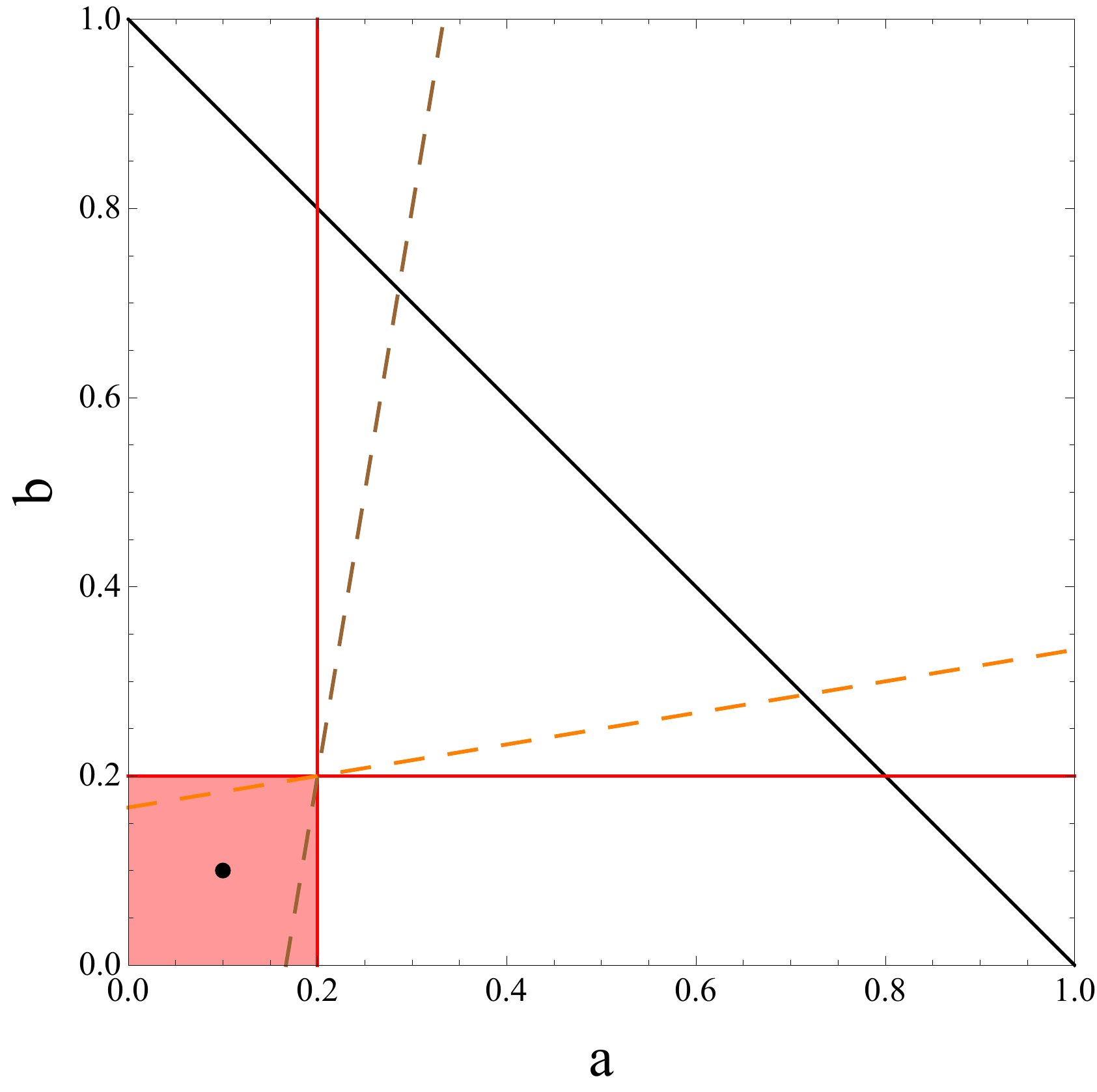} }     

  \caption{ Plot (a) shows the overlap of type-IV (also of type-II, III). Red point is inside the two-dimensional blue region and blue point is inside the two-dimensional red region. These are two possible conditions for type-IV overlap to exist. Plot (b) shows the red region in $(\za,\zb)$ plane where type-IV overlap can exist. The vertical and horizontal red lines are $\za=2 \Dd_{\phi}-1$ and $\zb=2 \Dd_{\phi}-1$ respectively. The dashed orange and dashed brown lines are $\zb = \frac{2 \Dd_{\phi} -1}{2 \Dd_{\phi}} (\za+1)$ and $\za = \frac{2 \Dd_{\phi} -1}{2 \Dd_{\phi}} (\zb+1)$ respectively. These four lines intersect at $(\za,\zb)=(2\Dd_\phi-1, 2\Dd_\phi-1)$. Part (a) is plotted for the  black point $(\za=0.1,\zb=0.1)$.} 
  \label{type-IV}
\end{figure}
It is clear from the figure \ref{type-IV}(a) that overlap of type-IV exists when either the red point lies inside two-dimensional blue region or the blue point lies inside two-dimensional red region. The first condition implies $k_{s-}^{\tt R} \geq \CQ_{\Dl_s} ( k_{s+}^{\tt R}, \za ,\zb ) \,\cap\, k_{s+}^{\tt R} \geq \CQ_{\Dd_s} ( k_{s-}^{\tt R}, \za ,\zb )$ while the second implies $k_{s-}^{\tt B} \geq \CQ_{\Dl_t} ( k_{s+}^{\tt B}, \za ,\zb )  \,\cap\,  k_{s+}^{\tt B} \leq \CQ_{\Dd_t} ( k_{s-}^{\tt B}, \za ,\zb ) $. 

In $(\za,\zb)$ plane it is easy to check that this translates to $ \big( \zb \leq \frac{2 \Dd_{\phi} -1}{2 \Dd_{\phi}} (\za+1) \cap \za \leq 2 \Dd_{\phi} -1 \big) \bigcup \big( \za \leq \frac{2 \Dd_{\phi} -1}{2 \Dd_{\phi}} (\zb+1) \cap \zb \leq 2 \Dd_{\phi} -1 \big) $ or equivalently, $\za \leq 2 \Dd_{\phi} -1 \cap \zb \leq 2 \Dd_{\phi} -1$. This region is shown in figure \ref{type-IV}(b). 

\subsection*{Existence of type-III overlap}

\begin{figure}[t]
    \centering
    \subfigure[]{\includegraphics[width=0.35\textwidth]{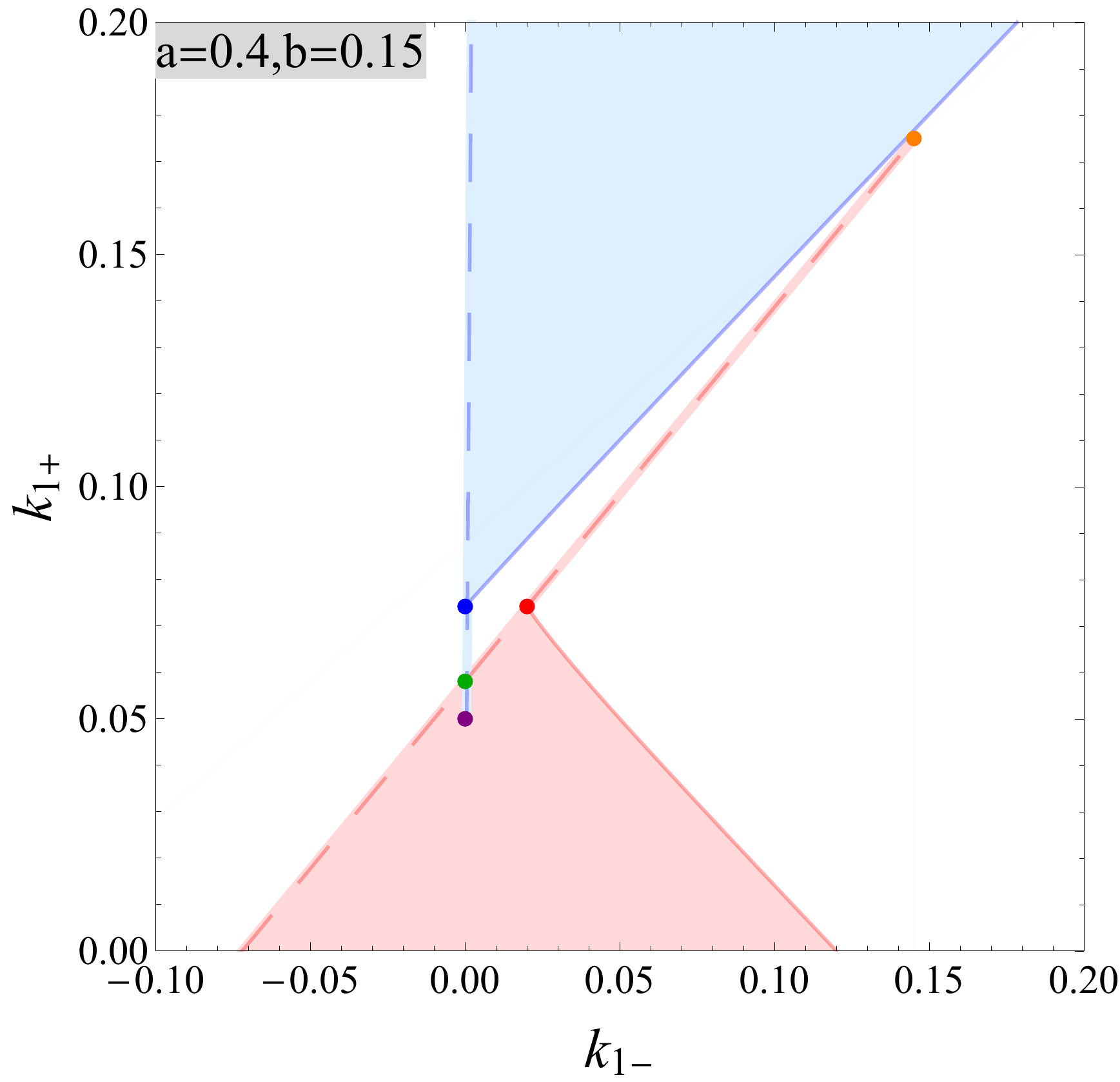} } \hspace{.5cm}
    \subfigure[]{\includegraphics[width=0.35\textwidth]{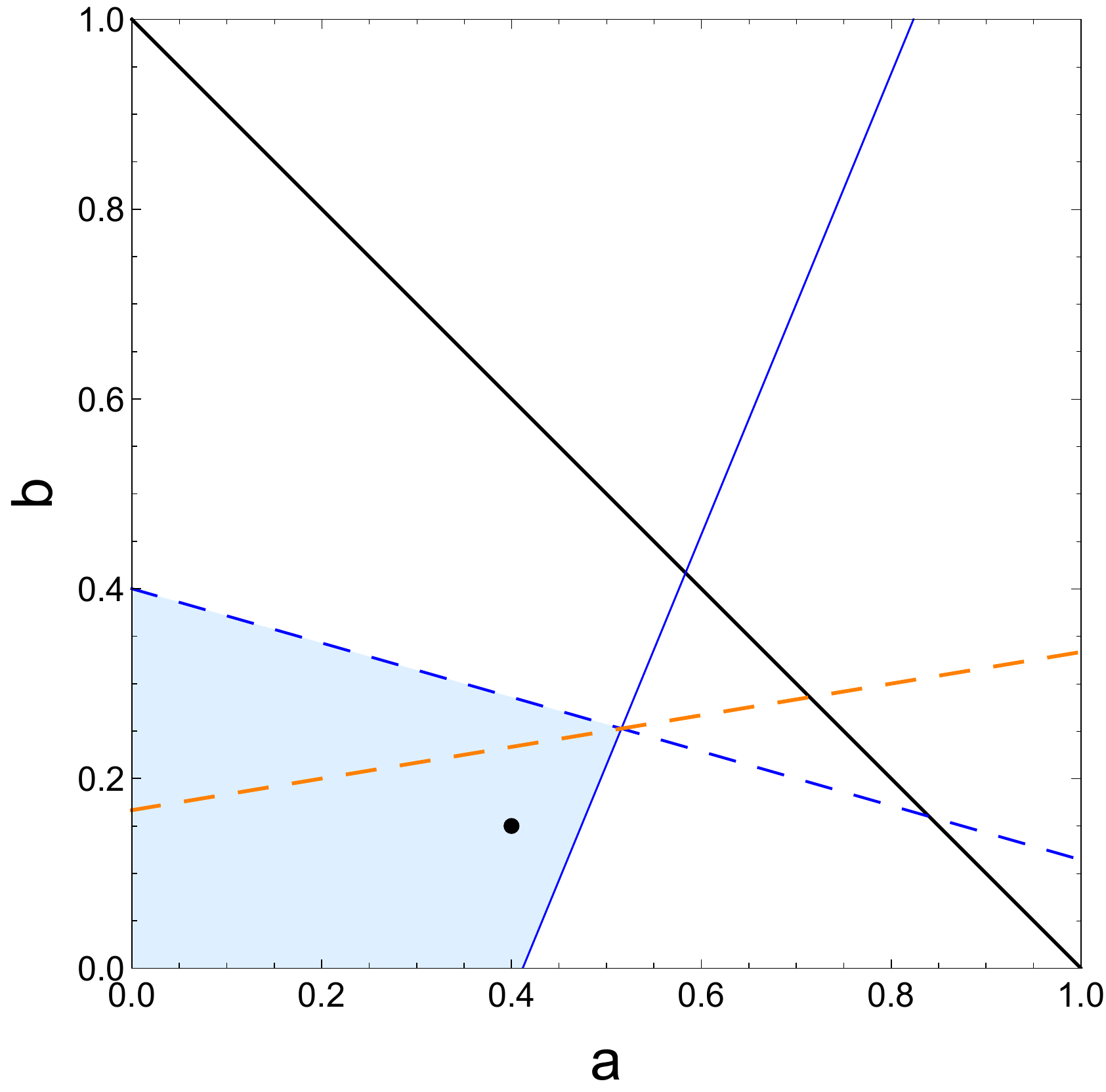} }     

  \caption{Plot (a) shows the overlap of type-III. Purple point  is inside the two-dimensional red region as required. Plot (b) shows the light-blue region in $(\za,\zb)$ plane where type-III overlap can exist. The blue and dashed blue lines are $\zb = \big( \frac{4\Dd_{\phi}+1}{4\Dd_{\phi}-1} \big) \za-1 $ and $ \zb = 2 \big( \frac{2\Dd_{\phi}-1}{4\Dd_{\phi}-1} \big) (4\Dd_{\phi}-\za-1) $ respectively. The dashed orange line is same as defined in figure \ref{type-IV}(b). All the three lines intersect at a point. The black point is $(\za=0.4,\zb=0.15)$ for which part (a) is plotted. } 
  \label{type-III}
\end{figure}
Overlap of type-III occurs when the purple point lies inside the red region as in  figure \ref{type-III}(a). This implies  $ \big( k_{s-}^{\tt P} \leq k_{s-}^{\tt R} \ \cap \ k_{s-}^{\tt P} \geq \CQ_{\Dl_t} ( k_{s+}^{\tt P}, \za ,\zb )  \big) \bigcup \big(   k_{s-}^{\tt P} \geq k_{s-}^{\tt R} \ \cap \ k_{s+}^{\tt P} \leq \CQ_{\Dd_t} ( k_{s-}^{\tt P}, \za ,\zb ) \  \big)  $. In $(\za,\zb)$ plane this translates to $\big(  \zb \leq \frac{2 \Dd_{\phi} -1}{2 \Dd_{\phi}} (\za+1) \ \cap \ \zb \geq \big( \frac{4\Dd_{\phi}+1}{4\Dd_{\phi}-1} \big) \za-1 \big)  \bigcup  \big( \ \zb \geq \frac{2 \Dd_{\phi} -1}{2 \Dd_{\phi}} (\za+1)\cap \zb \leq 2 \big( \frac{2\Dd_{\phi}-1}{4\Dd_{\phi}-1} \big) (4\Dd_{\phi}-\za-1)  \ \big)  $ or equally, $\zb \geq \big( \frac{4\Dd_{\phi}+1}{4\Dd_{\phi}-1} \big) \za-1 \cap \zb \leq 2 \big( \frac{2\Dd_{\phi}-1}{4\Dd_{\phi}-1} \big) (4\Dd_{\phi}-\za-1)     $, as shown in figure \ref{type-III}(b). \\

\subsection*{Existence of type-II overlap}
\begin{figure}[t]
    \centering
    \subfigure[]{\includegraphics[width=0.35\textwidth]{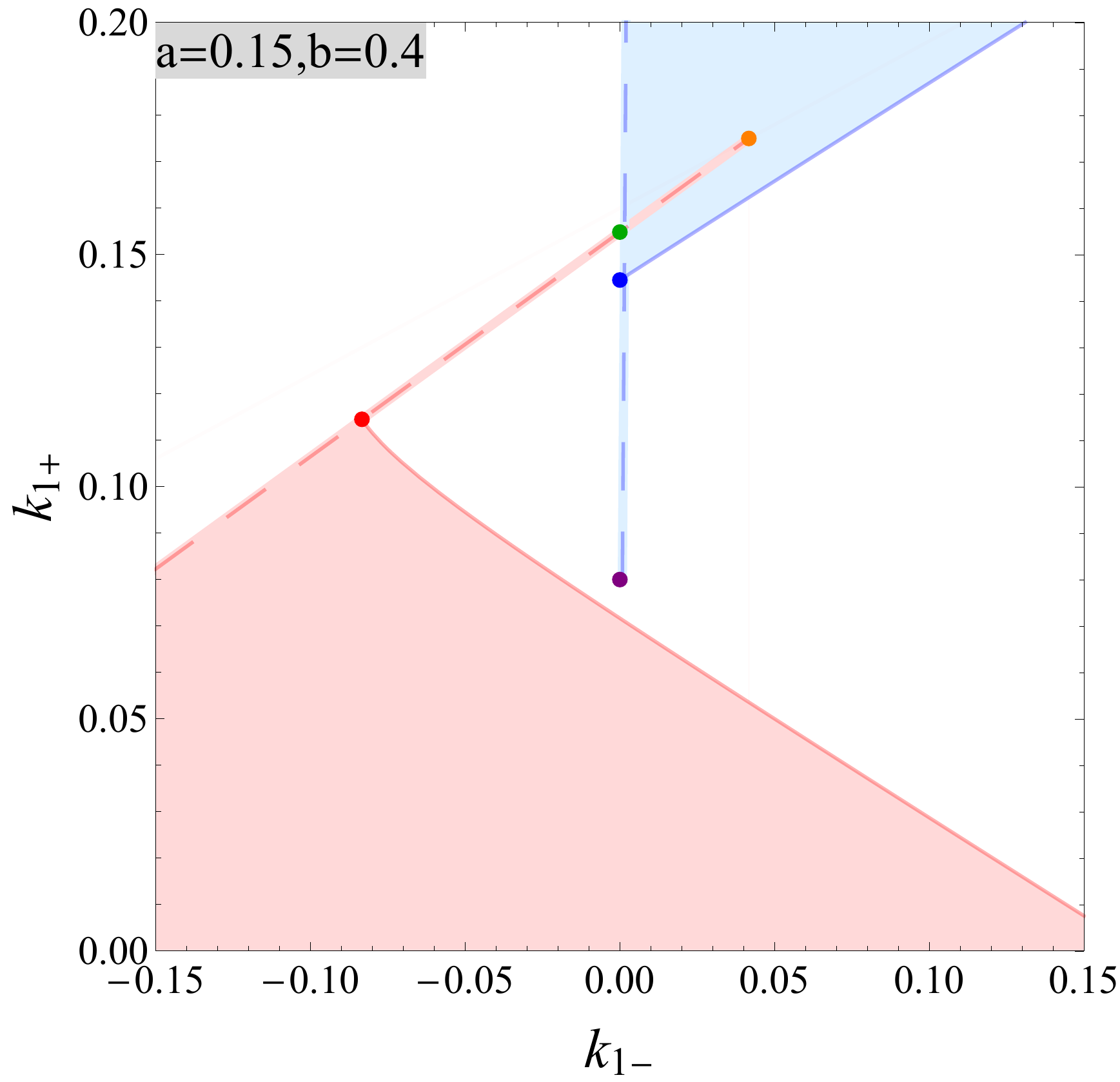} } \hspace{.5cm}
    \subfigure[]{\includegraphics[width=0.35\textwidth]{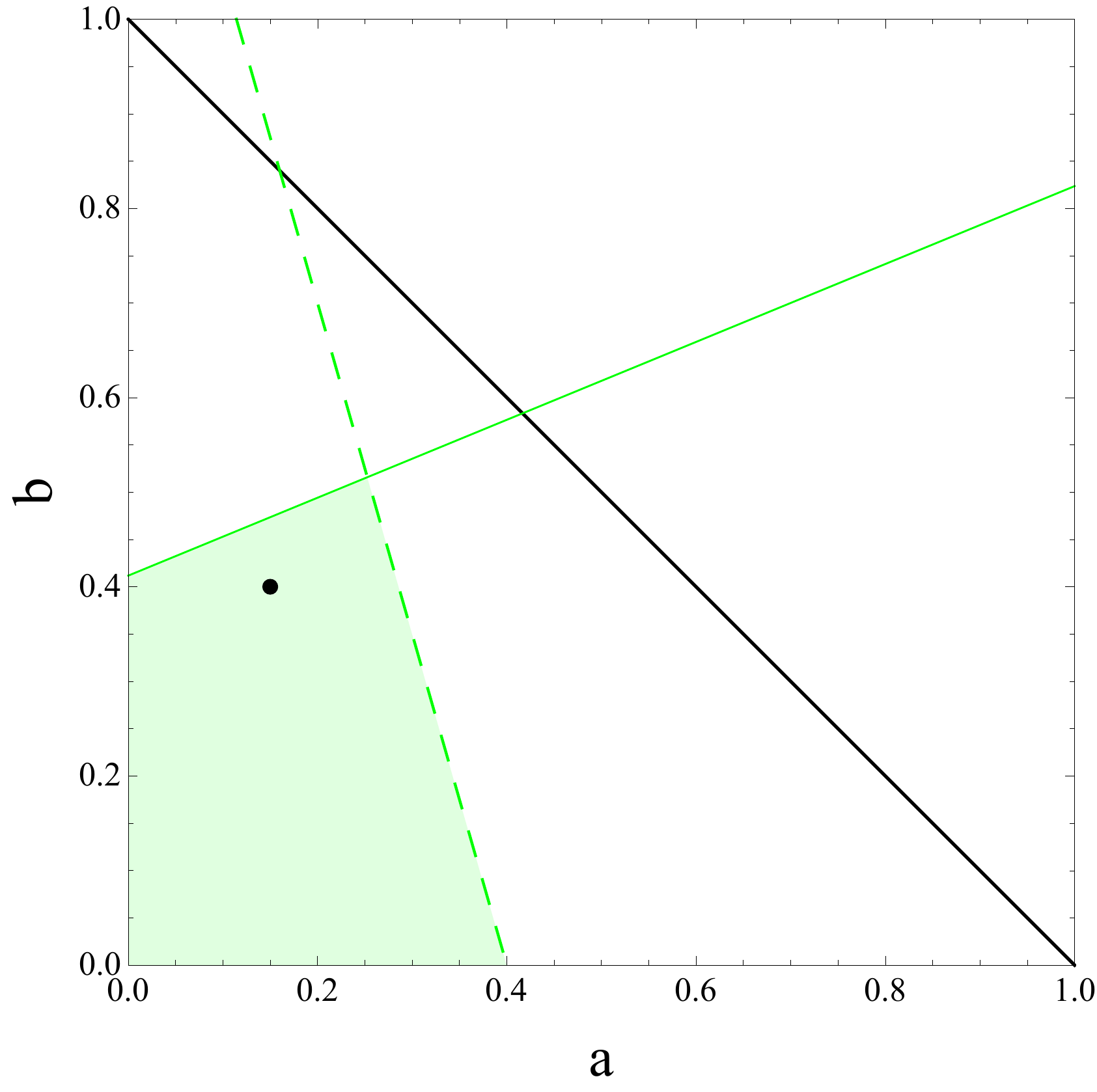} }     

  \caption{Plot (a) shows the overlap of type-II. Orange point is inside the two-dimensional blue region as required. Plot (b) shows the light-green region in $(\za,\zb)$ plane where type-II overlap can exist. The green and dashed green lines are $\za = \big( \frac{4\Dd_{\phi}+1}{4\Dd_{\phi}-1} \big) \zb-1 $ and $ \za = 2 \big( \frac{2\Dd_{\phi}-1}{4\Dd_{\phi}-1} \big) (4\Dd_{\phi}-\zb-1) $ respectively. The black point is $(\za=0.15,\zb=0.4)$ for which part (a) is plotted. } 
  \label{type-II}
\end{figure}
Type II is the dual of type II under crossing. It occurs when the orange point lies inside two-dimensional blue region as in the figure \ref{type-II}(a). This implies  $k_{s-}^{\tt O} \geq \CQ_{\Dl_s} ( k_{s+}^{\tt O}, \za ,\zb ) \cap k_{s+}^{\tt O} \geq \CQ_{\Dd_s} ( k_{s-}^{\tt O}, \za ,\zb )  $. In $(\za,\zb)$ plane this translates to $\za \geq \big( \frac{4\Dd_{\phi}+1}{4\Dd_{\phi}-1} \big) \zb-1 \cap \za \leq 2 \big( \frac{2\Dd_{\phi}-1}{4\Dd_{\phi}-1} \big) (4\Dd_{\phi}-\zb-1) $, as shown in figure \ref{type-II}(b). \\

\subsection*{Existence of type-I overlap}
\begin{figure}[t]
    \centering
    \subfigure[]{\includegraphics[width=0.35\textwidth]{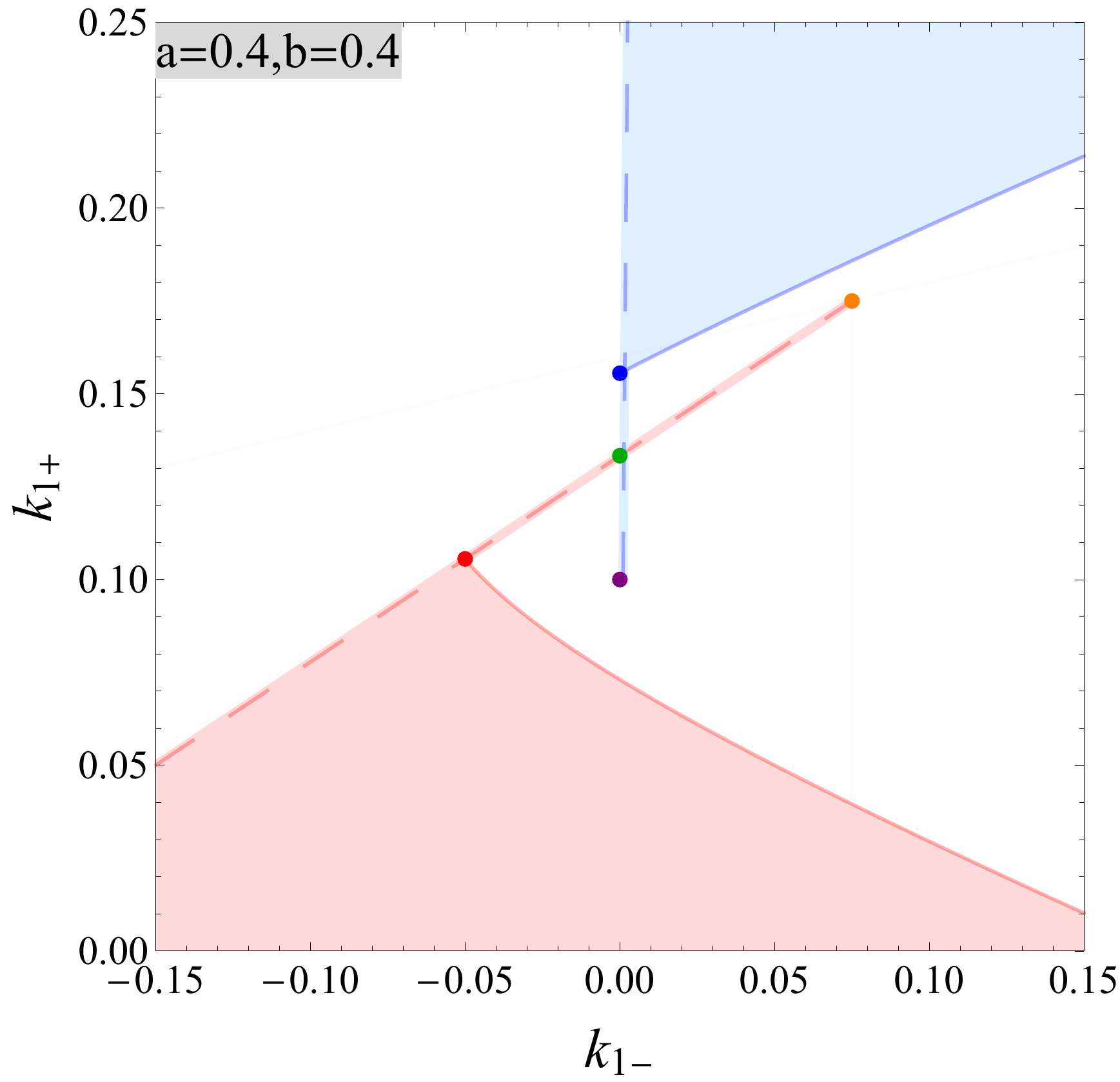} } \hspace{.5cm}
    \subfigure[]{\includegraphics[width=0.35\textwidth]{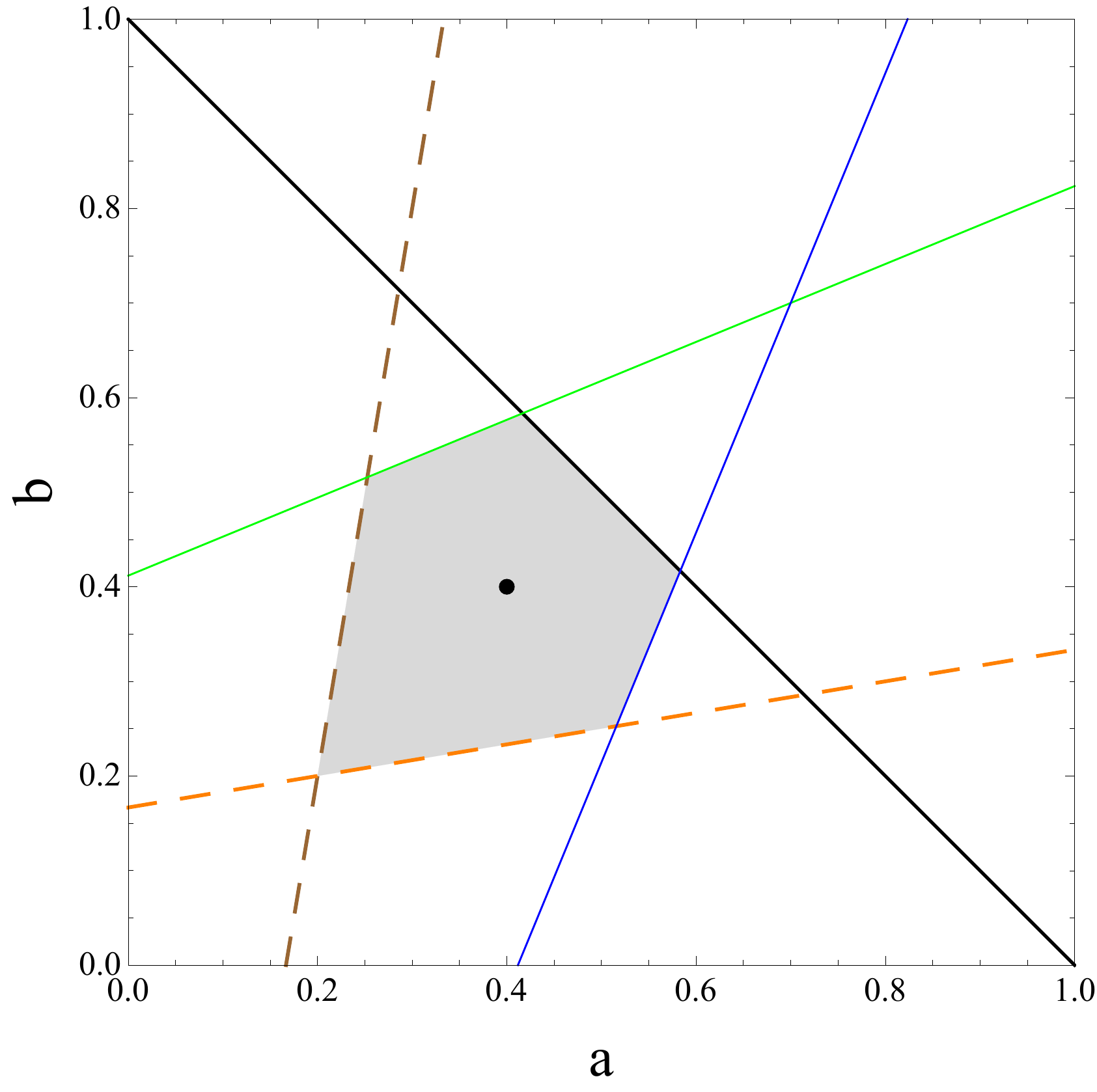} }

  \caption{ Plot (a) shows the overlap of type-I. Blue and red one dimensional regions intersect each other at green point as required. Plot (b) shows the gray region in $(\za,\zb)$ plane where type-I overlap can exist. The blue, green, dashed orange and dashed brown lines are the same ones as shown in the previous figures. The black point is $(\za=0.4,\zb=0.4)$ for which part (a) is plotted. } 
  \label{type-I}
\end{figure}
The overlap of type-I is simply the intersection of one-dimensional blue and one-dimensional red regions as shown in figure \ref{type-II}(a). In other words, it is the intersection of the curves $ k_{s-} = \CQ_{\Dl_s} ( k_{s+}, \za ,\zb ) $ and $ k_{s-} = \CQ_{\Dl_t} ( k_{s+}, \za ,\zb ) $ with $\Dd_s^*<1, \Dd_t^*<1$. This implies $ ( k_{s+}^{\tt G} \leq k_{s+}^{\tt B} \cap k_{s+}^{\tt G} \geq k_{s+}^{\tt B} ) \bigcap ( k_{s+}^{\tt G} \geq k_{s+}^{\tt R} \cap k_{s+}^{\tt G} \leq k_{s+}^{\tt O} )  $. In $(\za,\zb)$ plane this translates to $ \big( \zb \geq \big( \frac{4\Dd_{\phi}+1}{4\Dd_{\phi}-1} \big) \za-1 \ \cap \ \za  \geq  \big( \frac{2\Dd_{\phi}-1}{2\Dd_{\phi}} \big) (\zb+1) \big) \ \bigcap \ \big( \za \geq \big( \frac{4\Dd_{\phi}+1}{4\Dd_{\phi}-1} \big) \zb-1 \ \cap \ \zb  \geq    \big( \frac{2\Dd_{\phi}-1}{2\Dd_{\phi}} \big) (\za+1) \big) $, as shown in figure \ref{type-I}(b). \\

\begin{figure}[t]
    \centering
    \subfigure[No-overlap]{\includegraphics[width=0.31\textwidth]{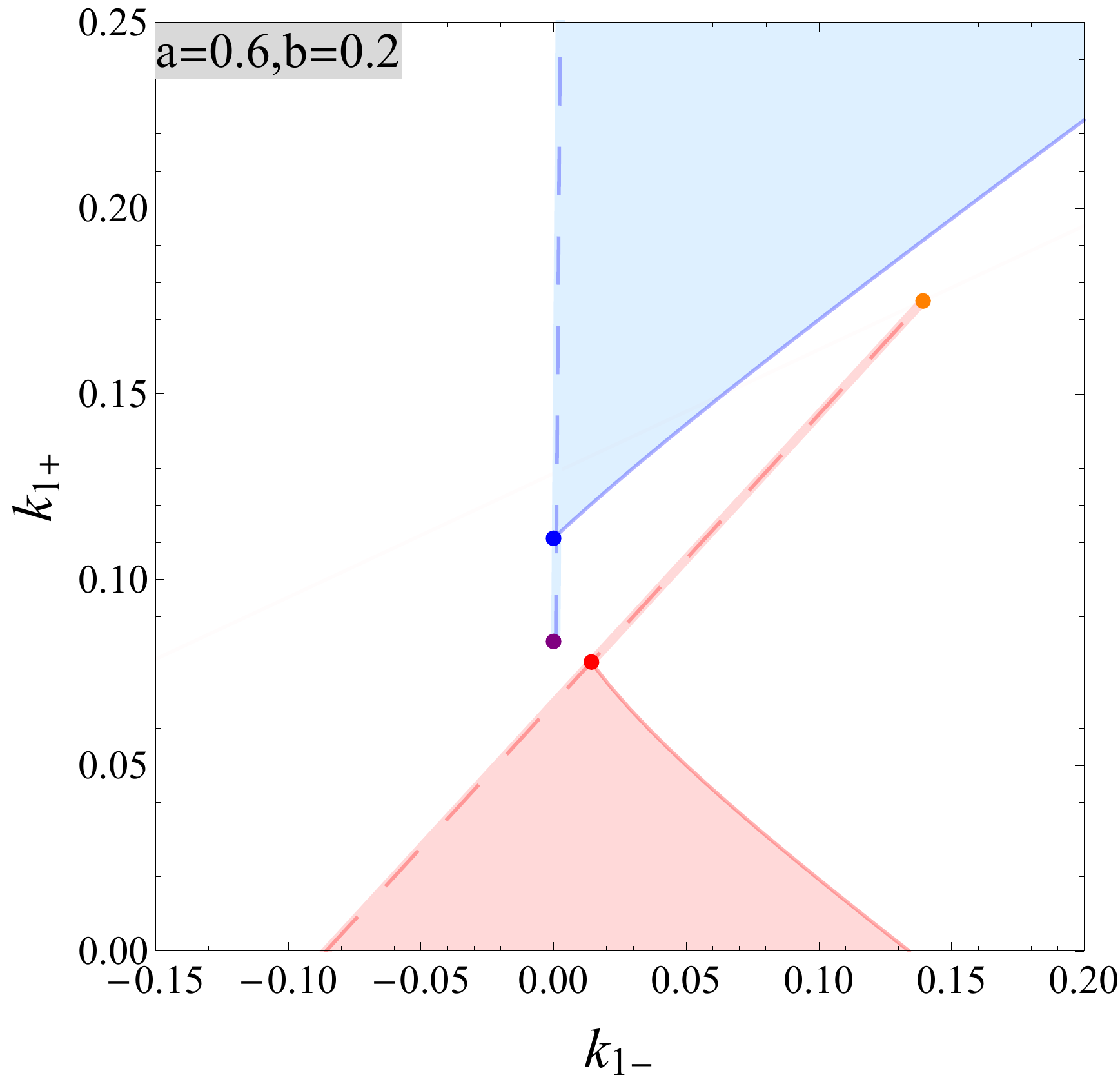} } \hspace{.1cm}
    \subfigure[Type-II and III]{\includegraphics[width=0.31\textwidth]{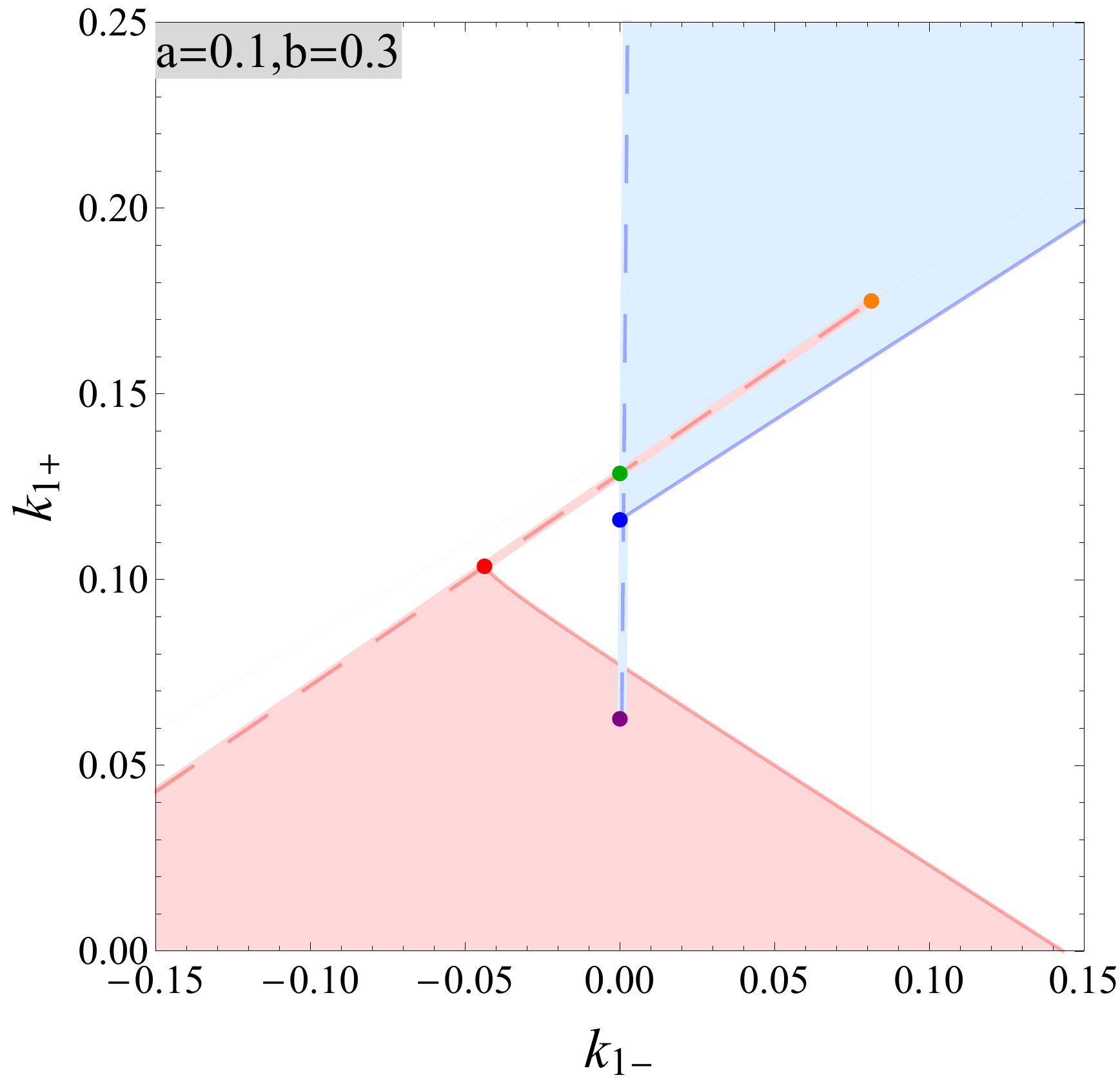} } 
\hspace{.1cm}
    \subfigure[Type-I, II and III]{\includegraphics[width=0.31\textwidth]{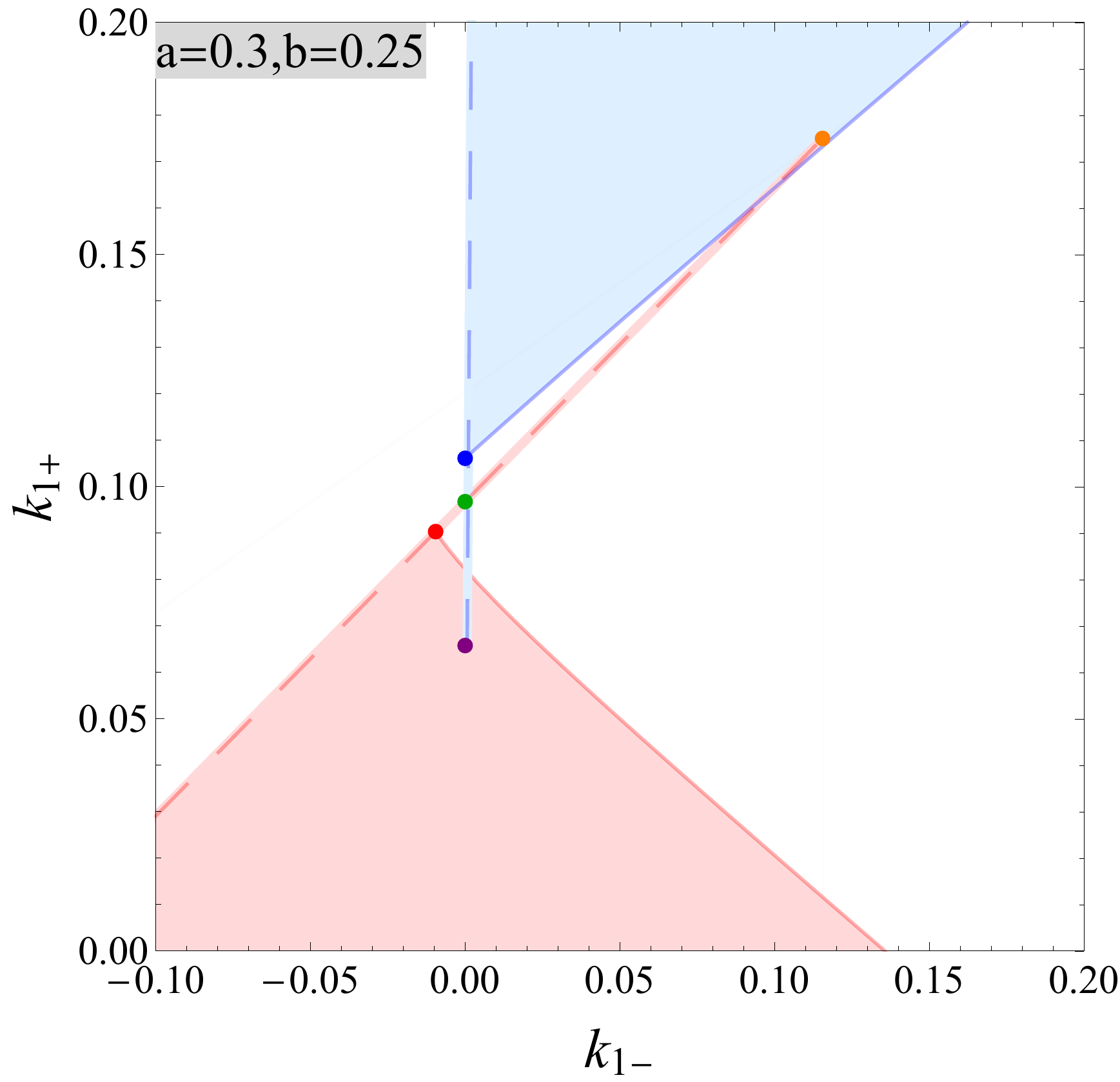} } 
      \caption{Some more examples of overlap diagrams.} 
  \label{overlap-examples}
\end{figure}

For some mixed types of overlaps see figure \ref{overlap-examples}. Here also we have taken $\Dd_{\phi}=0.6$.

\bibliographystyle{JHEP}
\bibliography{LargeDCFT}

\end{document}